\documentclass[twocolumn,showpacs,preprintnumbers,amsmath,amssymb,apsrev]{revtex4}

\usepackage{graphicx}
\usepackage{dcolumn}
\usepackage{bm}

\begin{document}

\title{Bubble statistics and coarsening dynamics for quasi-two dimensional foams with increasing liquid content}
\author{A. E. Roth, C. D. Jones, and D. J. Durian}
\affiliation{Department of Physics \& Astronomy, University of Pennsylvania, Philadelphia, PA 19104-6396, USA}

\date{\today}

\begin{abstract}
We report on the statistics of bubble size, topology, and shape and on their role in the coarsening dynamics for foams consisting of bubbles compressed between two parallel plates.  The design of the sample cell permits control of the liquid content, through a constant pressure condition set by the height of the foam above a liquid reservoir.  We find that in the scaling regime, all bubble distributions are independent not only of time but also of liquid content.   For coarsening, the average rate decreases with liquid content due to the blocking of gas diffusion by Plateau borders inflated with liquid; we achieve a factor of four reduction from the dry limit.  By observing the growth rate of individual bubbles, we find that von~Neumann's law becomes progressively violated with increasing wetness and with decreasing bubble size.  We successfully model this behavior by explicitly incorporating the border blocking effect into the von~Neumann argument.  Two dimensionless bubble shape parameters naturally arise, one of which is primarily responsible for the violation of von~Neumann's law for foams that are not perfectly dry.
\end{abstract}


\maketitle

\section{Introduction}

Coarsening is a process in foams where gas diffuses from one bubble to another, so that some bubbles grow and some bubbles shrink \cite{WeaireHutzlerBook}.  Coarsening also occurs elsewhere, such as for grains in metal alloys, and can often be treated by similar approaches \cite{GlazierWeaire92, Stavans93}.  Hence foam systems can be studied to understand coarsening behavior more generally.  This is simpler to accomplish in two dimensions, where bubble areas are readily measured by conventional digital imaging.  For ideal dry foams, which have zero liquid content and obey Plateau's rules, John von~Neumann \cite{VonNeumann} famously showed that the coarsening rate of a given bubble is exactly
\begin{equation}
\frac{{\rm d}A_{i}}{dt} = K_0 (n_{i}-6),
\label{VN}
\end{equation}
where $A_{i}$ and $n_i$ are respectively the area and number of sides of bubble $i$ (see Section~\ref{model} for a generalized derivation).  The constant of proportionality, $K_0$, is proportional to the film tension, the solubility and diffusivity of the gas in the liquid, and inversely proportional to the film thickness.  It is remarkable that neither the size nor shape of a bubble matters, only its number of sides.

There have been numerous experiments with dry two-dimensional foams to measure coarsening rates and other properties, such as area and side number distribution functions.  This includes direct measurements on dry soap froths \cite{GlazierGrossStavans87, GlazierStavans89, StavansGlazier89, Stavans90, Stavans93sf, Icaza94}, soap froths with different boundary conditions \cite{StavansKrichevsky92, RosaFortes99, RosaFortes02}, and measurements on lipid monolayers \cite{KnoblerPRA90, Bergeetal90}.  Simulations have also been performed \cite{KermodeWeaire90, GlazierAndersonGrest90, HerdtleAref92, Segeletal93, NeuSch97, Rutenberg05}.  This body of work shows good general agreement with von~Neumann's law.

While von~Neumann's law describes the rate of change of area for individual bubbles in dry two-dimensional foam, it also bears on how the average bubble area, $\langle A \rangle = \sum_{i} A_{i} / N_{total} = A_{total}/N_{total}$, changes with time.  Following the argument of Ref.~\cite{Stavans93}, first note that the average square bubble area, $\langle A^{2} \rangle = \sum_{i} {A_{i}}^{2} / N_{total}$, depends on the width of the area distribution and hence would seem to depend on foam production method and coarsening history.  But in fact coarsening foams tend to evolve into a self-similar growth regime, where distribution shapes are stationary and do not depend on time except for an overall scale factor.  Once this scaling regime is reached, the quantity $\langle A \rangle ^{2} / \langle A^{2} \rangle$ is constant.  Therefore the identity
\begin{equation}
\frac{\langle A^{2} \rangle}{\langle A \rangle ^{2}} \langle A \rangle = \frac{1}{A_{total}} \sum_{i=1}^{N_{total}} {A_{i}}^{2}
\end{equation}
may be differentiated with respect to time, from ${\rm d} \langle A\rangle /{\rm d}t$ on the left and from ${\rm d} A_i^2/{\rm d}t = 2A_i{\rm d}A_i/{\rm d}t = 2A_iK_0(n_i-6)$ on the right.  The result can be rearranged and expressed as follows,
\begin{align}
\frac{d \langle A \rangle}{dt} &= 2K_0 \frac{\langle A \rangle ^{2}}{\langle A^{2} \rangle} \sum_{n} F(n) (n-6), \\
&= 2K_0 \frac{\langle A \rangle ^{2}}{\langle A^{2} \rangle} [\langle \langle n \rangle \rangle - 6],
\label{avgareaeqn}
\end{align}
by introducing a new quantity, the area-weighted side-number distribution
\begin{equation}
F(n) = \sum_{i~s.t.~n_{i} = n} A_{i} / A_{total}.
\label{fndef}
\end{equation}
By this definition $F(n)$ represents the probability that a randomly chosen point in space lies inside an $n$-sided bubble, which is distinct from the widely-studied probability $p(n)$ that a randomly chosen bubble is $n$-sided.  In the scaling regime, according to Eq.~(\ref{avgareaeqn}), the rate of change of average bubble area depends on the shape of the area distribution via $\langle A \rangle ^{2} / \langle A^{2} \rangle$ and the area-weighted average number of sides per bubble, defined in Eq.~(\ref{avgareaeqn}) by $\langle \langle n \rangle \rangle = \sum_{n} n F(n)$.  The distribution $F(n)$, and in particular the difference of $\langle \langle n \rangle \rangle$ from 6, thus play an important role in the evolution of the foam.  However, we are unaware of previous experimental or theoretical investigation of area-weighted statistical quantities, by contrast with numerous studies of $p(n)$.

Coarsening in three dimensional foams has also been extensively studied, and the generalization of von~Neumann's law is now known~\cite{MacPhersonSrolovitz2007}.  In terms of experiment, most studies of coarsening in three dimensions have been on wet foams.  Various techniques include multiple light scattering \cite{DurianWeitzPine90, DurianWeitzPine91a, DurianWeitzPine91b, HutzlerWeaire00}, magnetic resonance imaging \cite{Gonatusetal95}, optical tomography \cite{Adler98}, x-ray tomography \cite{GlazierGraner05, GlazierGraner10}, and observation of surface bubbles \cite{Jameson99, StoneKoehlerHilgenfeldt01, VeraDurian02, Marquezetal04, FeitosaDurian08}.  However it is much easier to work experimentally with two dimensional foams, where individual bubbles are readily imaged.

One aspect of coarsening that has not been fully elucidated is the effect of non-zero liquid fraction, $\varepsilon$.  Experiments on this effect have primarily focused on coarsening rates of three dimensional foams.  One study suggested a mechanism for the reduced coarsening rate of three dimensional wet foams as the reduced film area due to liquid in the Plateau borders covering regions of the films and measured under forced drainage that the coarsening rate was reduced by a factor of $(1-\sqrt{\varepsilon / 0.36})$ \cite{HutzlerWeaire00}.  Another study measured coarsening rates for a freely draining three dimensional foam and using this model of Plateau border blocking film area measured that the coarsening rate was reduced by a factor of $(1-\sqrt{\varepsilon / 0.44})^{2}$ \cite{StoneKoehlerHilgenfeldt01}.  Other studies on coarsening in three dimensional wet foams have found empirically that the coarsening rate is reduced by a factor of $1/\sqrt{\varepsilon}$ \cite{VeraDurian02, FeitosaDurian08}.  In two dimensions, there has been theoretical \cite{BoltonWeaire91, Weaire99, Fortes05, Mancini07} and simulation \cite{BoltonWeaire91, BoltonWeaire92} work on the effects of liquid fraction on coarsening.  And while this paper was in preparation, a new theoretical approach was proposed, and tested by Potts model simulations, based on an effective number of sides that depends on the fraction of the perimeter occupied by wet versus dry interfaces \cite{Graneretalarxiv}.  Ref.~\cite{Marchalot08} describes coarsening experiments on bubbles in a microfluidic geometry, where there is a non-zero liquid content that affects the growth rate of average bubble area and that is modeled by an average effective film permeability.  Despite all this activity, we are unaware of any work that systematically measures or models the bubble-level topology-dependent effects of liquid content on coarsening.

To make progress on these issues, we present a series of experiments in which the liquid content is systematically varied and the size, shape, and topology of individual bubbles are measured as a function of time.  We begin with a description of the foaming system, the sample cell, and the imaging techniques.  After demonstrating the success of these methods, we report on bubble statistics, which turn out all to be independent of both time and liquid content.  Then we consider the coarsening rate, how it varies with liquid content, and how it develops a violation of von~Neumann's law.  Finally we present a model to quantitatively explain this behavior.

\section{Materials and Methods}

\begin{figure}
\includegraphics[width = 3.00in]{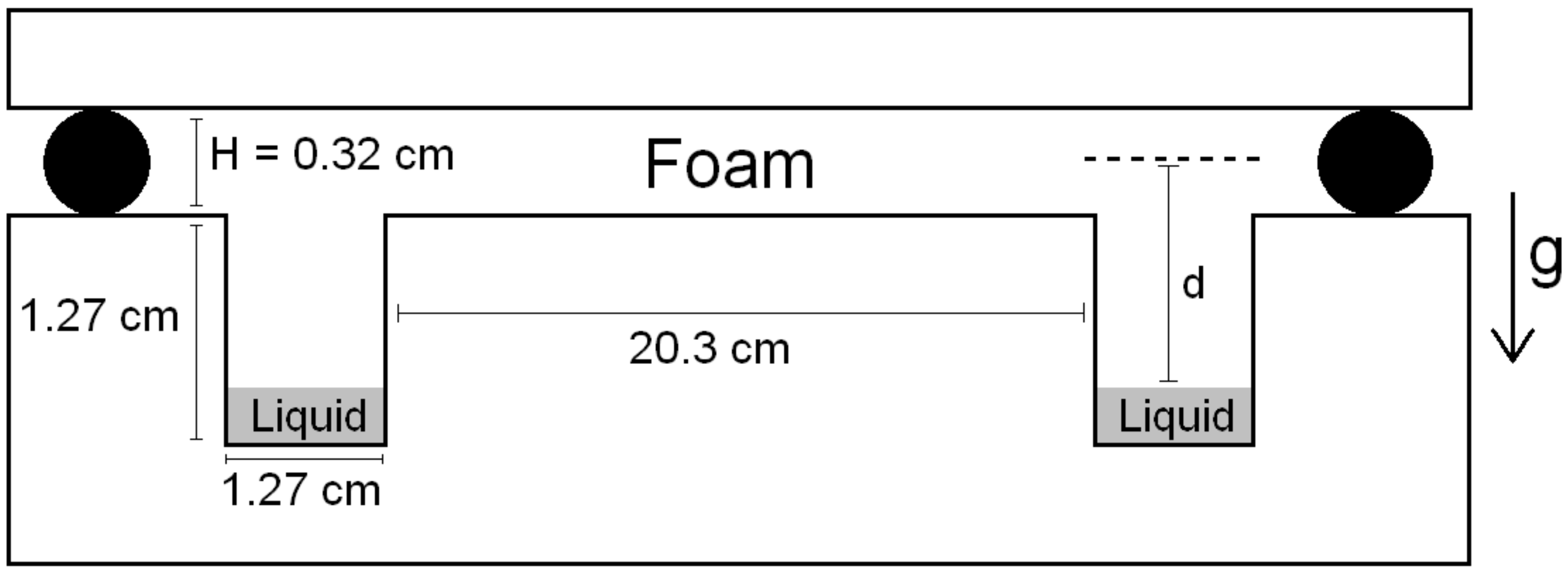}
\caption{A schematic cross section of the circular constant pressure cell; not to scale.  Measurements are made in a central $11 \times 11$ cm$^{2}$ region of interest.  The foam wetness is controlled by filling depth of liquid in the trough, in terms of Eq.~(\ref{rvdequation}) and the distance $d$ of the top of the liquid reservoir below the center of the foam.  The two solid black circles represent a cross section of the inner O-ring.  Not shown: outer O-ring, O-ring grooves, two filling ports, bolt circle between the two O-rings, spacers.}
\label{schematic}
\end{figure}

The liquid foaming solution consists of 75\% deionized water, 20\% glycerin, and 5\% Dawn Ultra Concentrated dish detergent, and has liquid-vapor surface tension $\gamma = 25~{\rm dynes} / {\rm cm}$.  This creates foams that are stable and long lived; film ruptures were never observed.  The sample cell consists of a circular chamber made from clear 1.91~cm thick acrylic plates separated by a $H=3.2$~mm gap and sealed with two concentric rubber O-rings, the inner of which is 23~cm in diameter.  The gap thickness and seal are maintained by a bolt circle and metal spacers, all between the two O-rings.  A cross section of  the cell is schematically in Fig.~\ref{schematic}.  To create the foam, the chamber is first completely filled with solution.  Pure nitrogen is then pumped into the chamber until only the desired amount of liquid remains.  This is accomplished via two valved ports attached on opposite sides of the bottom plate.  The chamber is then shaken vigorously until it is completely filled by a uniform opaque foam with sub-millimeter size bubbles, smaller than the gap between the plates.  The initial foam is thus three dimensional.  Immediately after production it is placed 20~cm away from a Vista Point A lightbox, and 2.5~m from a Nikon D80 camera with a Nikkor AF-S 300 mm 1:2.8 D lens.  It is then left undisturbed to coarsen into a two-dimensional foam consisting of a single layer of bubbles with an average size greater than the gap, which typically requires two days.  The field of view thus encompasses up to a few hundred bubbles.  Under computer control, photographs are then taken at two-minute intervals for durations ranging up to two weeks.  From all runs, a total of 14663 bubbles were observed.  This is enough for statistical purposes, though it is possible to observe many more bubbles at lower resolution using sample cells that are larger or have thinner gaps \cite{Quilliet2008, Duplat2011}.

The crucial innovative feature of the sample cell is a circular trough, of width and depth 1.27~cm and inner diameter 20.3~cm, which serves as both a liquid reservoir and a means to control the liquid content of the foam.  The initial three-dimensional foam is quite wet, but it drains by gravity and the expelled liquid accumulates in the trough.  As the foam becomes drier, the radius of curvature $r$ of the Plateau borders decreases and the Laplace pressure $\gamma/r$ increases.  Drainage halts when hydrostatic equilibrium is established by balance of capillary and gravity forces.  For this, the Laplace pressure must equal the gravitational pressure $\rho g d$, where $\rho=1.07$~g/cc is the liquid solution density, $g=980$~cm/s$^2$, and $d$ is the distance of the Plateau borders above the liquid in the reservoir, as depicted in Fig.~\ref{schematic}.  Accordingly, the radius of curvature of the Plateau borders is given by 
\begin{equation}
r = \frac{\gamma}{\rho g d},
\label{rvdequation}
\end{equation}
and hence can be controlled through $d$ by the filling depth of liquid in the reservoir.  Here $d$ is measured to $\pm0.2$~mm and the dimensions of the reservoir trough are large enough that this depth remains constant once the foam becomes two dimensional.  Thus the coarsening of interest proceeds at constant, controllable, $r$.  A further advantage of the trough is that the relatively large volume of liquid solution permits easy foam production by shaking.

\begin{figure*}
\includegraphics[width = 6.00in]{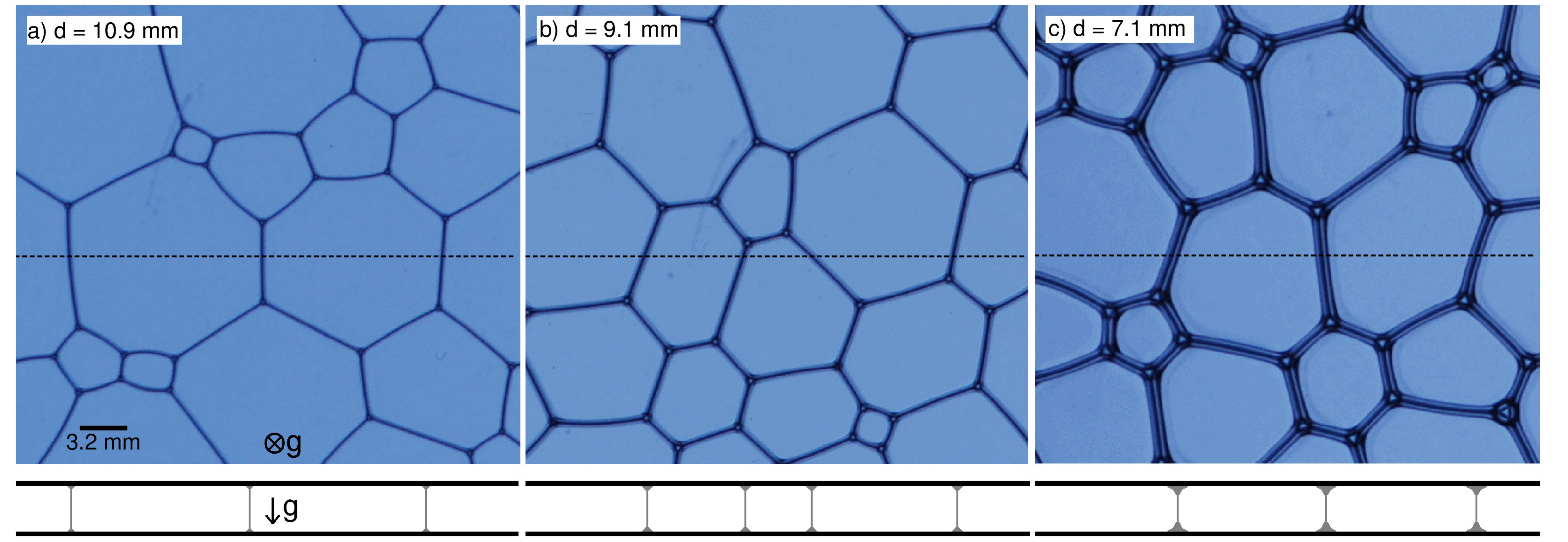}
\caption{(Color online) Images of a subregion of three foams, with different liquid filling depths $d$ as labelled. For smaller $d$, the Plateau border radius $r$ increases according to Eq.~(\ref{rvdequation}).  This is evident in the main images and is shown underneath by schematic drawings of surface Plateau borders and soap films in a vertical cross-section along the dotted lines in the middle of the main images.  The scale for all images and schematics is indicated by the bar in (a), which equals the gap $H$ between top and bottom plates and hence the height of the vertical soap films.  In (c) note that there are bright spots at the `vertices' where three surface Plateau borders are seen to meet.  This feature arises from light channeled up through the thick vertical Plateau borders that span the gap between the upper and lower plates of the sample cell.  Note that these are well separated; therefore, there is no ambiguity in determining the number $n$ of sides of  a bubble, even in the wettest foams measured here.}
\label{threelfs}
\end{figure*}

Example images are shown in Fig.~\ref{threelfs} for foams with three different filling depths, $d$, which decrease from left to right.  It can be seen that as $d$ decreases, the Plateau borders become noticeably thicker as expected by Eq.~(\ref{rvdequation}).  While the foams appear to be dry and two-dimensional, their actual three-dimensional structure is emphasized underneath the main images by schematic drawings of a vertical cut across each foam.  There, the Plateau borders running along the top and bottom plates appear as scalloped triangular regions, and the soap films running between plates appear as vertical lines connecting top and bottom Plateau borders.  Bubble area is thus appropriately measured by the skeletonization procedure as the area enclosed by the vertical soap films, not as the ``free area'' seen by eye to be enclosed by thick Plateau borders.  Note that variation of $d$ affects only the Plateau borders, not the film thicknesses.  Since the Plateau borders are macroscopic, while the film thickness is of order 100~nm, the liquid content of the foam is set entirely by the Plateau border thickness.  The volumetric liquid fraction scales as $r^2 R / (R^2 H) \propto 1/(d^2 R)$ where $R$ is the typical bubble radius and $H$ is the gap between the plates.  The projected-area liquid fraction scales as $r R/ (R^2) \propto 1/(d R)$.  Neither of these liquid fractions remains constant as the foam coarsens; rather, more importantly, the Plateau border radii and Laplace pressures remain constant as set by the distance $d$ of the foam above the top of the liquid reservoir.  Throughout, we thus refer to $d$ as controlling the liquid {\it content}, not the liquid fraction.

\begin{table}[ht]
\begin{ruledtabular}
\begin{tabular}{ccc}
$d$ (mm)&$N_{initial}$&$N_{final}$\\
\colrule
11.3& 114 & 41 \\
10.9& 73 & 18\\
9.4& 144 & 44\\
9.1& 298 & 143\\
8.5& 104 & 82\\
8.0& 384 & 49\\
7.1& 290 & 158\\
6.7& 217 & 85\\
6.2& 252 & 100\\
\end{tabular}
\end{ruledtabular}
\caption{\label{DepthsAndNumbers} Initial and final numbers of bubbles in an $11 \times 11~{\rm cm}^{2}$ square region of interest in the center of the cell for different liquid filling depths.  The quantity $d$ is the distance of the foam above the liquid reservoir.  The uncertainty in $d$ is 0.2~mm.  Only bubbles completely within the region of interest are considered.}
\end{table}

Digital images such as shown in Fig.~\ref{threelfs} are collected for foams with a wide range of different filling depths, as listed in Table~\ref{DepthsAndNumbers} along with the number of bubbles entirely in the central $11\times11$~cm$^2$ region of interest at the beginning and end of the collection period.  Using standard procedures, it is relatively straightforward to threshold and skeletonize each image and then measure the area and number of sides of each bubble that lies entirely within the region of interest.  However, when a small bubble shrinks toward zero its diameter inevitably becomes smaller than the distance between the plates.  Then it may `pinch in' and form a film horizontally in the middle of the bubble, and thus no longer be two-dimensional.  Such bubbles and their neighbors, are excluded from the analysis.

\begin{figure*}
\includegraphics[width = 6.00in]{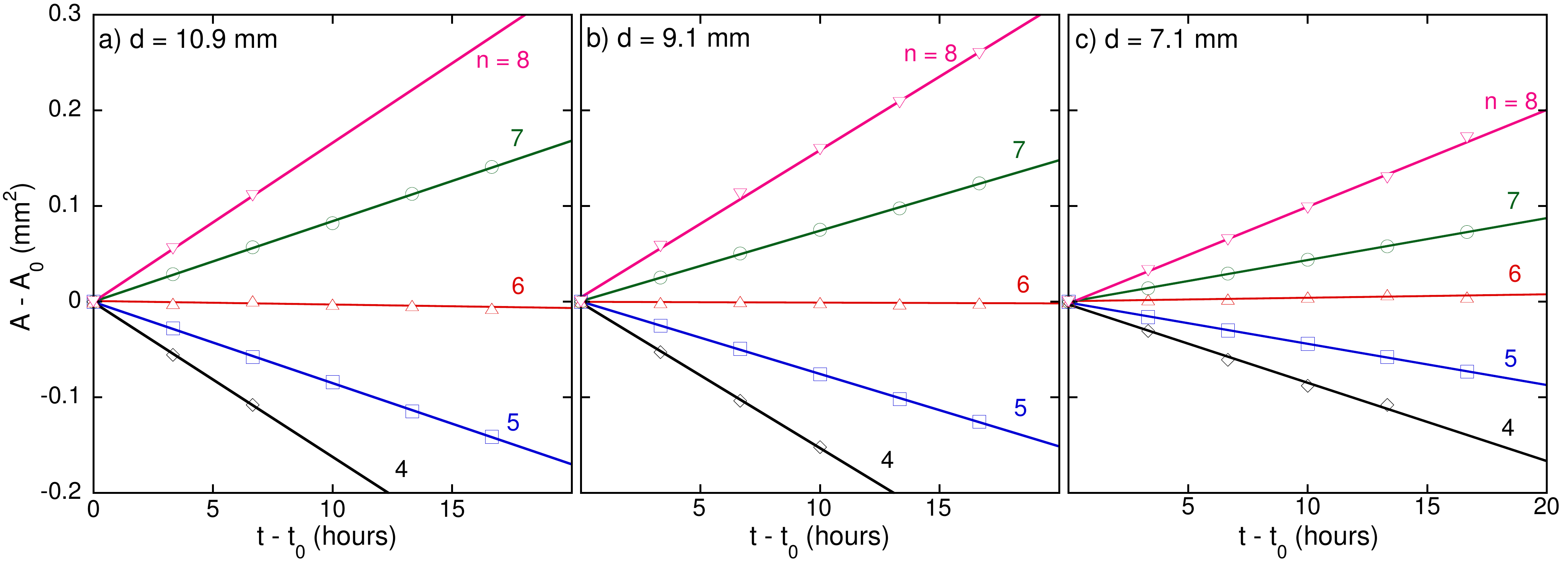}
\caption{(Color online) Area versus time for selected bubbles with different number $n$ of sides, for three different distances $d$ of the foam above the reservoir, as labelled.  The liquid content increases with decreasing $d$, as illustrated in Fig.~\ref{threelfs}.   Note that the area change is linear in time and at rate proportional to $(n-6)$.  The lines are fits to the von~Neumann form, $A-A_0 = K (n-6)t$, with the same $K$ value for all $n$:
(a) $K=0.84\pm0.06~{\rm mm}^{2} / {\rm hr}$, (b) $K=0.72\pm0.06~{\rm mm}^{2} / {\rm hr}$, (c) $K=0.42\pm0.02~{\rm mm}^{2} / {\rm hr}$,
Increasing the liquid content decreases the rate of change of area, such that wetter foams coarsen more slowly; compare to Fig.~\ref{coarseningvarea}.}
\label{avgavt}
\end{figure*}

Example results for area versus time are displayed in Fig.~\ref{avgavt} for individually selected bubbles with different side numbers $n$, for the  same three foams depicted above with different liquid filling depths.  Note that the areas are constant for $n=0$, and either increase or decrease linearly with time for $n>6$ or $n<6$, respectively.  Fits are found to the von~Neumann prediction, $A(t)=A_0+K(n-6)$, where $A_0$ is the area at an initial time and a single value of $K$ is adjusted to simultaneously fit all the data in each panel of the figure.   While these fits are excellent, the feature of main interest in Fig.~\ref{avgavt} is that {\it the coarsening rate decreases with increasing liquid content}, as $d$ decreases from left to right.  Indeed the slopes for a given $n$ are equal to $K(n-6)$ and are seen to decrease by a factor of two from (a) to (c).  Intuitively, the thicker the Plateau border, the smaller the film area through which gas diffuses, and hence the slower the coarsening.  This serves as proof-of-principle:  Our custom sample cell design and procedures thus succeed in producing dry two-dimensional foams with controllable Plateau border thicknesses.

As a technical aside, throughout the remainder of the paper the rate ${\rm d}A/{\rm d}t$ of a bubble's growth is found by fits of $A(t)$ vs $t$ over a time window over which the side number $n$ remains constant.  And there is no ambiguity in the value of $n$, even for the wettest foams at smallest $d$ values where $r$ becomes as large as $H/4$, since the foams have large enough bubbles to appear two-dimensional when viewed from above, as in Fig.~\ref{threelfs}.  In other words, the soap films remain vertical and are easily located by the thresholding/skeletonization procedure for any wetness.  Even in the wet foam limit, where horizontal top and bottom Plateau borders merge, the vertical Plateau borders are still well-separated and hence $n$ is well-defined.  The only difficulty is for very small three-sided bubbles, which can detach from the top or bottom plate and hence become three-dimensional.  Since three-sided bubbles tend to start small and shrink rapidly, they do not  remain two-dimensional for very long.  Due to this effect, we were able to measure growth rates for only eight of the 195 three-sided bubbles seen in our combined runs.

\begin{figure}
\includegraphics[width=3.00in]{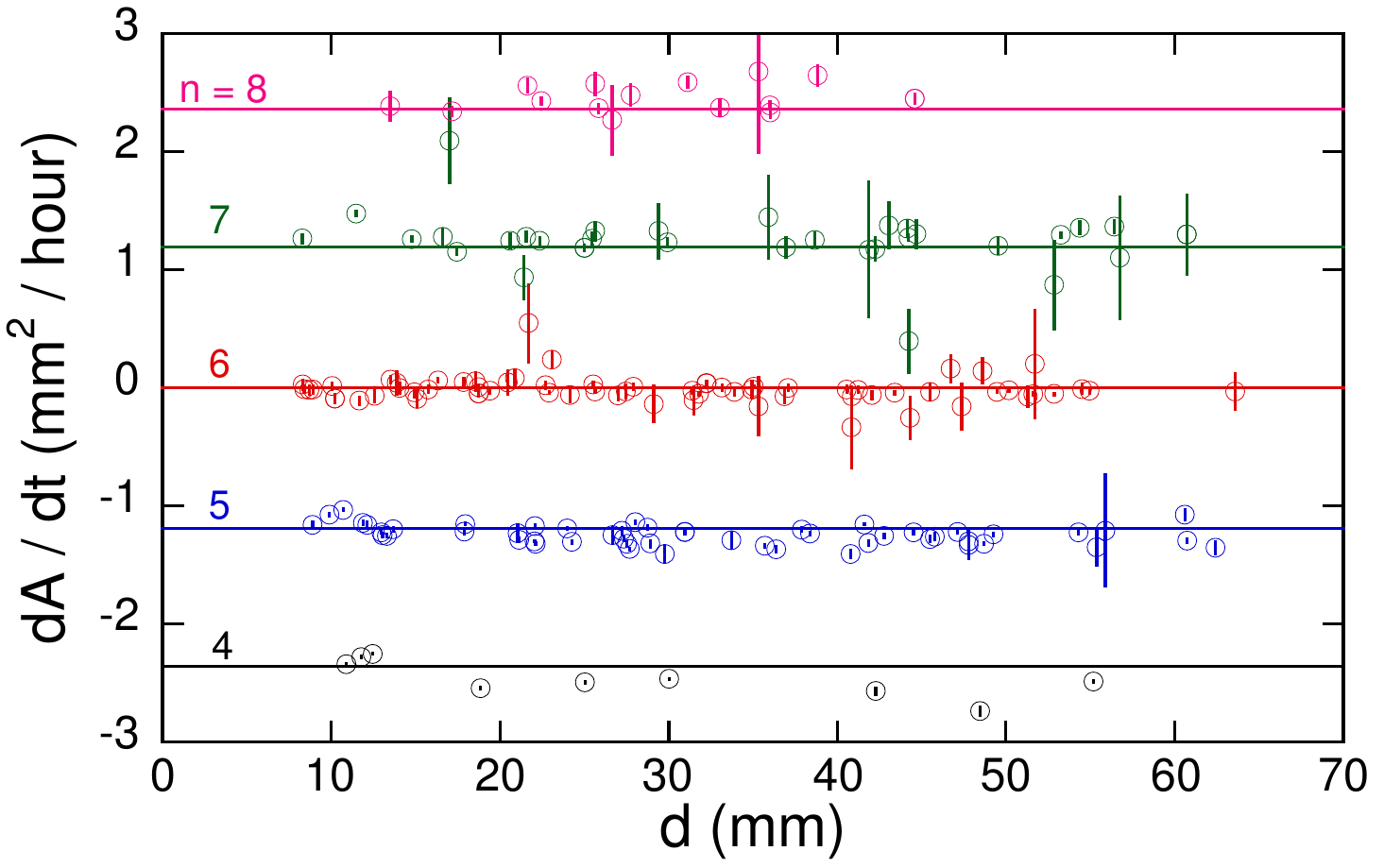}
\caption{(Color online) Rates of area change for bubbles in a vertical cell vs height $d$ of bubbles above the liquid surface.  The cell is the same shown in Fig.~\ref{schematic}, but re-oriented and filled 7.5~cm from the bottom of the O-ring.  Symbol types distinguish bubbles with different number $n$ of sides, as labeled.  The lines represent ${\rm d}A / {\rm d}t = K_0(n-6)$ (Eq.~(\ref{VN})) with $K_0 = 1.20\pm0.06~{\rm mm}^{2}/{\rm hr}$ .  Since the growth rates are independent of $d$, the bubbles are in the dry foam limit where the Plateau border size is negligible compared to bubble size.}
\label{vertical}
\end{figure}

To further characterize our liquid solution, we now measure coarsening in the very dry limit where the border thickness is made as small as possible.  For this, we use the same sample cell but orient it vertically rather than horizontally and fill it with liquid to a depth of 7.5~cm above the bottom of the O-ring.  As usual, foam is produced by vigorous shaking and then allowing it to drain and coarsen for about one day into a two-dimensional froth.  The rate of area change, ${\rm d}A/{\rm d}t$ of individual bubbles is then measured along with their number of sides and their height $d$ above the drained liquid.  Since the cell is vertical, the value of $d$ can be up to 6~cm, which is much greater than can be attained in the horizontal orientation due to the fixed 1.27~cm depth of the trough.   By Eq.~(\ref{rvdequation}), this gives the smallest Plateau border radius as 0.005~cm.  The resulting coarsening rates are plotted vs $d$ in Fig.~\ref{vertical}, with each point representing one bubble with side numbers indicated by symbol color and label.  Note that ${\rm d}A/{\rm d}t$ depends on side number but has no apparent dependence on $d$ across the entire range of $1~{\rm cm}<d<6$~cm.  These data are therefore all in the dry foam limit.  Furthermore, absence of dependence on $d$ indicates that the film thickness is constant.  In principle the thickness must decrease with height due to gravity, but apparently a balancing disjoining pressure can be achieved by very slight thinning away from the minimum in the effective interface potential.   The fit to von~Neumann's law, ${\rm d}A/{\rm d}t = K_0(n-6)$, is shown by the solid horizontal lines, and gives $K_0=1.20\pm0.06$~mm$^2$/hr.  This value reflects the physical chemistry of the gas/surfactant-solution/soap-film system, independent of the geometry of the bubbles and the Plateau borders.

This completes the description of materials and methods, and the characterization of the foaming system.  In the next sections we now turn to the main tasks of measuring bubble statistics and coarsening rates as a systematic function of liquid content.

\section{Bubble Statistics}

In the follow three subsections we present the statistical distributions for the topology, size, and shapes of bubbles. 

\subsection{Topology}

\begin{figure}
\includegraphics[width = 3.00in]{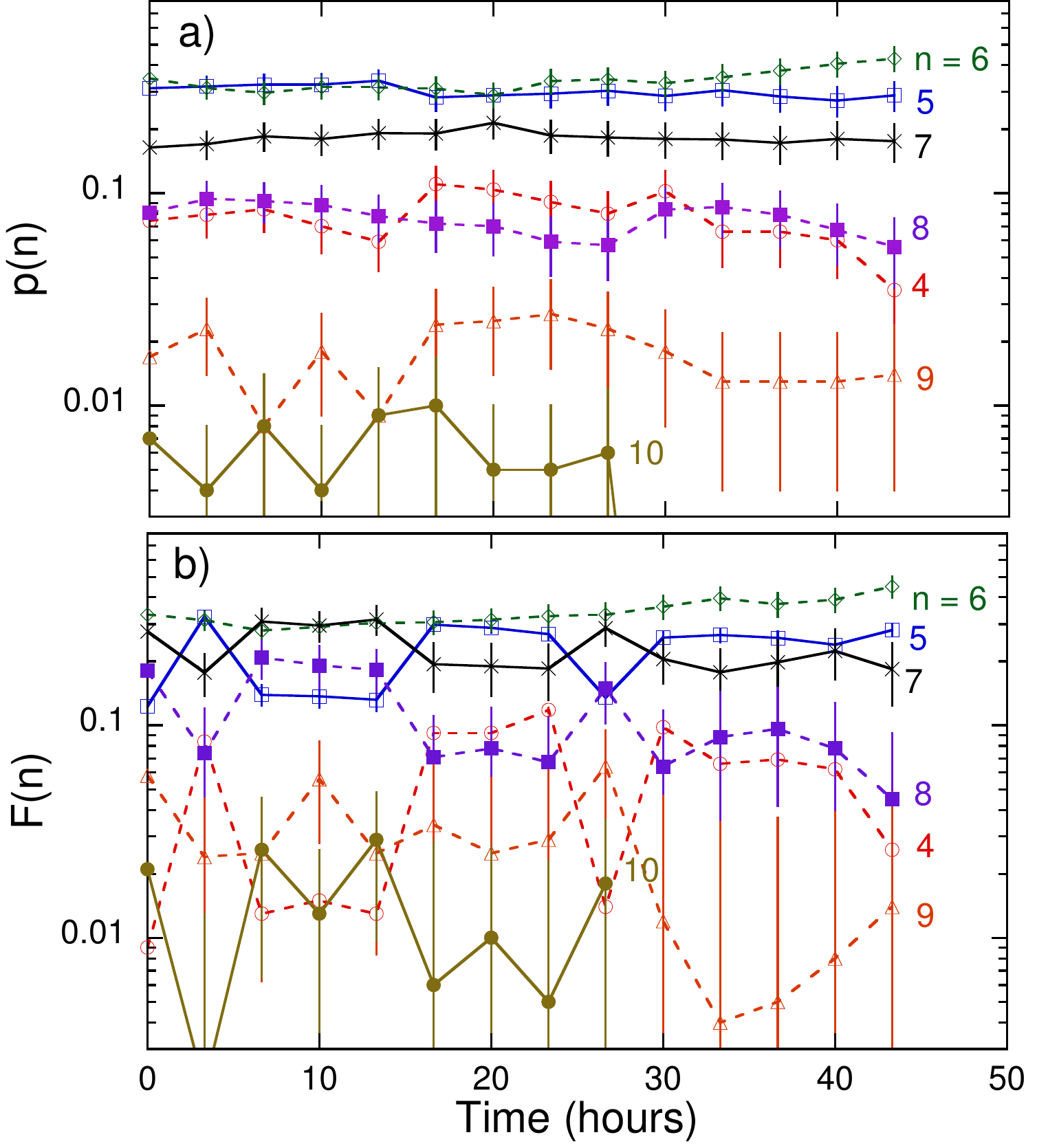}
\caption{(Color online) (a) Side number distribution, and (b) area-weighted side number distribution, versus time for a typical foam sample with $d = 9.1$~mm.}
\label{pvt}
\end{figure}

\begin{figure}
\includegraphics[width = 3.00in]{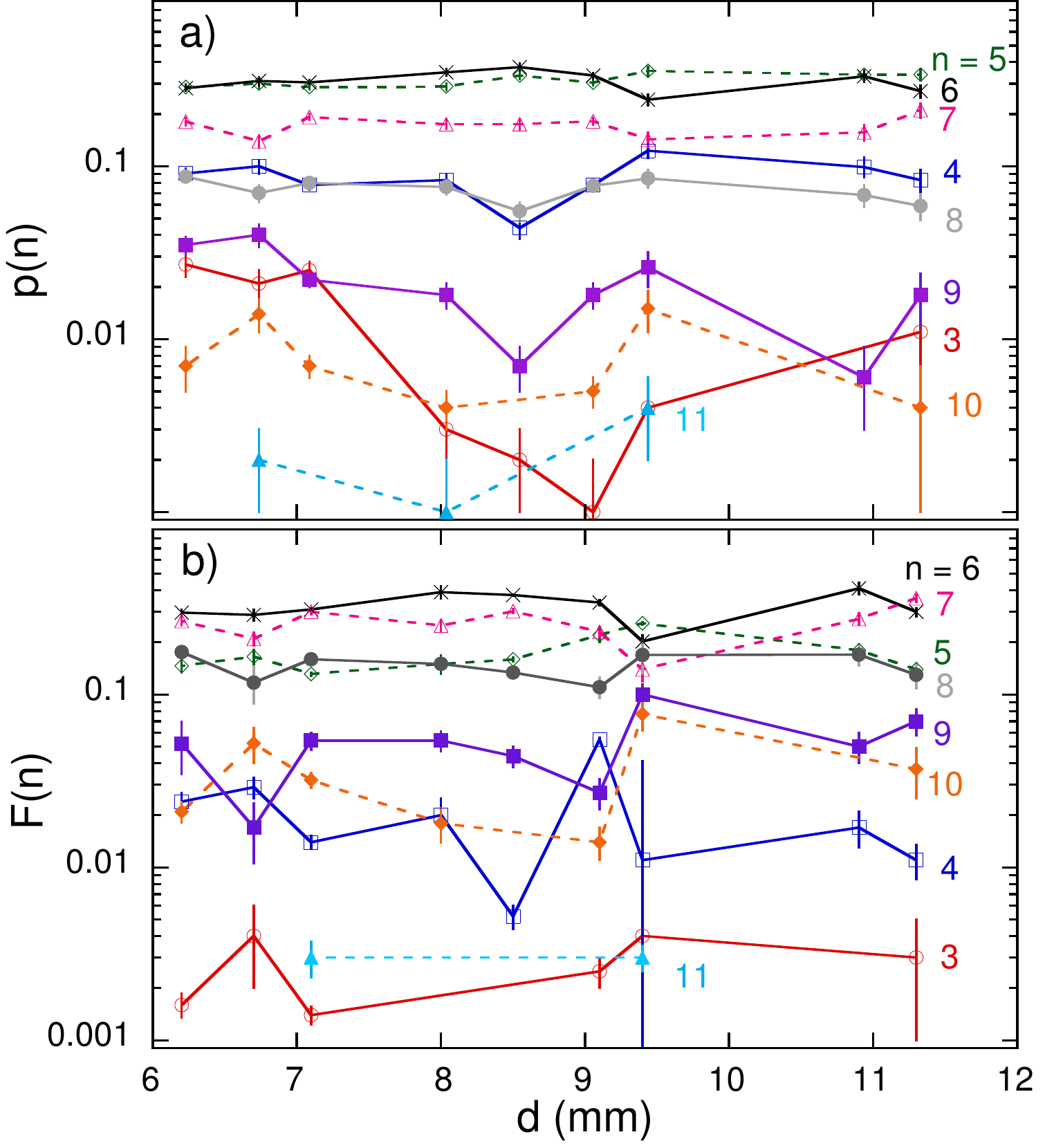}
\caption{(Color online) Time averages of (a) side number distribution, and (b) area-weighted side number distribution, versus height $d$ of the foam above the liquid reservoir.}
\label{plf}
\end{figure}

\begin{figure}
\includegraphics[width = 3.00in]{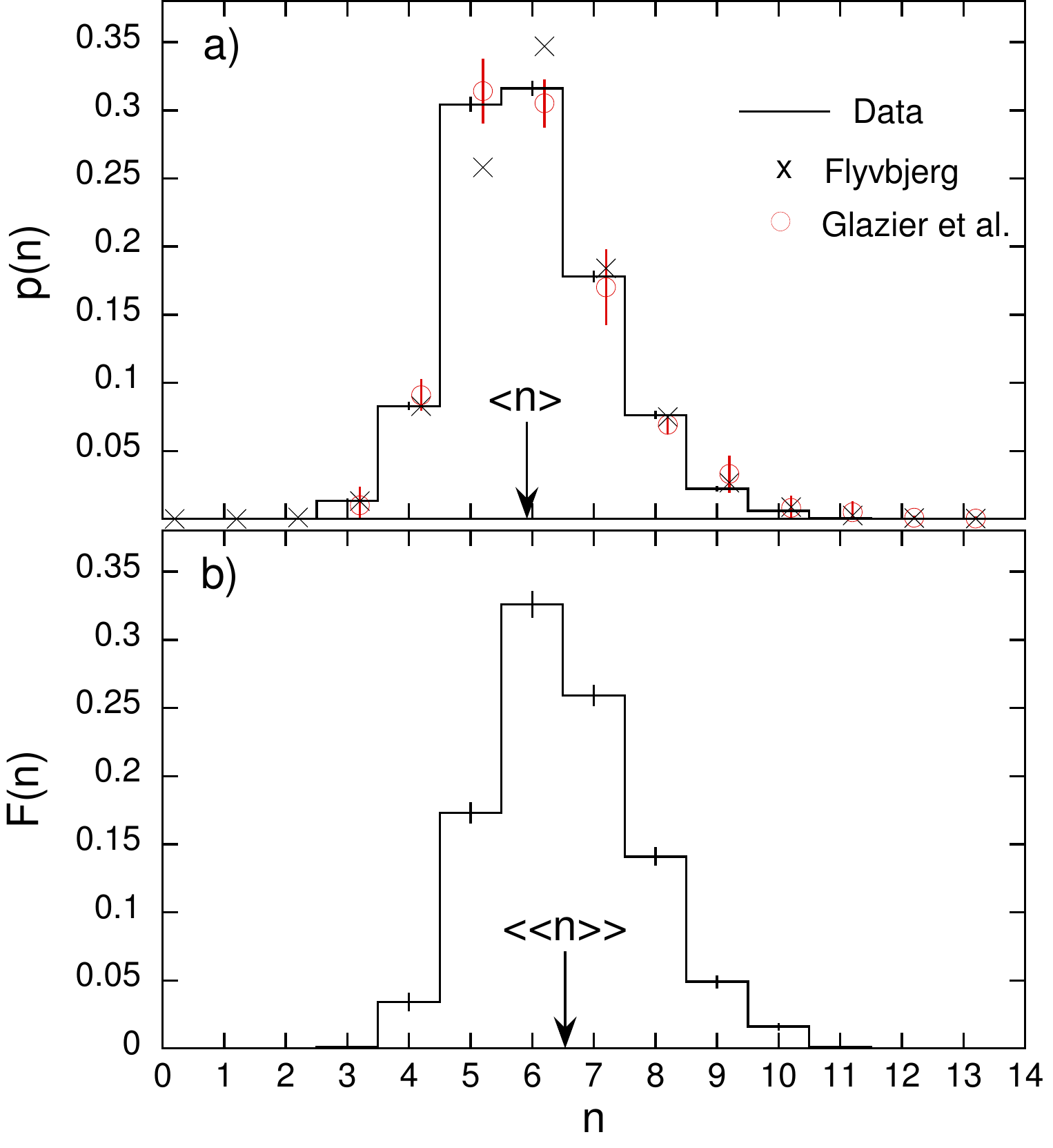}
\caption{(Color online) (a) Side number distribution, and (b) area-weighted side number distribution, averaged over all times and over all liquid contents.  The average number of sides, and the area-weighted number of sides, are indicated by arrows as labeled.  For comparison with $p(n)$, data from Ref.~\cite{GlazierAndersonGrest90} are shown by open circles and the predictions from Ref.~\cite{Flyvbjerg93} are shown by crosses.  The plotted distribution data are listed in Table~\ref{distributiondata}.}
\label{pnstep}
\end{figure}

\begin{table*}[width = 3.00in]
\begin{ruledtabular}
\begin{tabular}{ccccc}
$n$ & $N$ & $p(n)$ & $F(n)$ & $m(n)$\\
\colrule
3 & 195 & $0.013 \pm 0.001$ & $0.0009 \pm 0.00006$ & $7.69 \pm 0.05$ \\
4 & 1217 & $0.083 \pm 0.002$ & $0.034 \pm 0.006$ & $7.04 \pm 0.02$ \\
5 & 4462 & $0.304 \pm 0.005$ & $0.173 \pm 0.007$ & $6.5 \pm 0.007$ \\
6 & 4634 & $0.316 \pm 0.005$ & $0.326 \pm 0.009$ & $6.22 \pm 0.006$ \\
7 & 2611 & $0.178 \pm 0.003$ & $0.259 \pm 0.007$ & $6.06 \pm 0.007$ \\
8 & 1120 & $0.076 \pm 0.002$ & $0.141 \pm 0.006$ & $5.92 \pm 0.01$ \\
9 & 327 & $0.022 \pm 0.001$ & $0.049  \pm 0.004$ & $5.82 \pm 0.02$ \\
10& 89 & $0.006 \pm 0.0006$ & $0.016 \pm 0.002$ & $5.73 \pm 0.03$ \\
11& 8 & $0.0005 \pm 0.0002$ & $0.001 \pm 0.0005$ & $5.91 \pm 0.16$ \\
\end{tabular}
\end{ruledtabular}
\caption{\label{distributiondata} Topological distributions averaged over all times and liquid contents, and their uncertainties.  Here $n$ is the number of sides; $N$ is the total number of bubbles observed with $n$ sides; $p(n)$ is the fraction of bubbles having $n$ sides, and the uncertainty is the value divided by $\sqrt{N}$; $F(n)$ is the fraction of area occupied by $n$ sided bubbles,  and the uncertainty is the standard deviation divided by the square root of the number of photographs; and $m(n)$ is the average number of sides of the neighbors of an $n$ sided bubble, and the uncertainty is the standard deviation divided by $\sqrt{N}$.  The total number of bubbles observed is $\sum N = 14663$.}
\end{table*}

The number of sides of a bubble is a key topological quantity, not just for describing the bubble but also for determining its coarsening rate according to von~Neumann's law.  Thus we begin by analyzing image data for the probability $p(n)$ that a randomly-chosen bubble has $n$ sides and also for the probability $F(n)$ that a randomly-chosen point in space is inside an $n$-sided bubble.  As discussed in the introduction, $F(n)$ is an area-weighted side number distribution that sets the average coarsening rate in the scaling regime.  Example data for these side number distributions are plotted, separately for each $n$, versus time in Fig.~\ref{pvt} for a typical foam sample with $d=9.1$~mm.  To within statistical uncertainty, the individual $p(n)$ and $F(n)$ values are seen to be independent of time.  This demonstrates that the foam is in a scaling regime, which is not surprising because the production method gave very small bubbles that coarsened greatly before data collection commenced.  This holds for the other foams with different liquid content, too, and therefore we may compute the time-averages of the side distributions.  The results for $p(n)$ and $F(n)$ are shown in Fig.~\ref{plf} versus the height $d$ of the foam above the liquid reservoir.  Now we see that, to within statistical uncertainty, there is no systematic dependence on liquid content.  This is consistent with the validity of the decoration theorem, as expected since vertical Plateau borders do not merge~\cite{BoltonWeaire91}.

Since the side distribution $p(n)$ and the area-weighted side distribution $F(n)$ do not vary with time or liquid content, we therefore average together all the data and plot the final results versus $n$ in Fig.~\ref{pnstep}.  Actual numerical values and uncertainties are given in Table~\ref{distributiondata}.  Both distributions are peaked at $n=6$ sides, and have full-width at half-maximum of about three.  Out of 14663 total bubbles, we never observed any with fewer than $n=3$ sides or with more than $n=11$ sides.   The detailed shape of $p(n)$ is consistent with prior observations  \cite{Stavans93, Stavans93sf, GlazierWeaire92, GlazierAndersonGrest90}, as shown by comparison with the data from Ref.~\cite{GlazierAndersonGrest90} and the theoretical prediction from Ref.~\cite{Flyvbjerg93}.  The shape of $F(n)$ is skewed from $p(n)$ toward higher $n$, which is expected because bubbles with larger $n$ tend to have greater area (as discussed in detail in the next sub-section).  To our knowledge, there is no prior data or theory with which to compare our $F(n)$ data.

\begin{table}[ht]
\begin{ruledtabular}
\begin{tabular}{ccc}
Quantity&Definition&Value\\
\colrule
$\langle n \rangle$& $\sum n p(n)$ & $5.92 \pm 0.01$\\
$\mu_{2}$ & $\sum [n - \langle n \rangle]^{2} p(n)$ & $1.56 \pm 0.02$\\
$\langle \langle n \rangle \rangle$ & $\sum n F(n)$ & $6.53 \pm 0.08$\\
$\nu_{2}$ & $\sum [n - \langle \langle n \rangle \rangle]^{2} F(n)$ & $1.67 \pm 0.09$\\
$\langle A^2 \rangle / \langle A \rangle^2$ & $[\sum A_{i}^2 / N_{tot}] / [\sum A_{i} / N_{tot}]^2$ & $1.72 \pm 0.25$\\
$\langle P^2 \rangle / \langle P \rangle^2$ & $[\sum P_{i}^2 / N_{tot}] / [\sum P_{i} / N_{tot}]^2$ & $ 1.20 \pm 0.06 $\\
\end{tabular}
\end{ruledtabular}
\caption{\label{bubbleaverages} Measured values of several statistical quantities, averaged over all times and liquid contents, and their uncertainties.  Here $n$ is the number of sides of a bubble; $p(n)$ is the fraction of bubbles with $n$ sides; $F(n)$ is the fraction of area occupied by $n$ sided bubbles; $A$ is bubble area; $P$ is bubble perimeter; and $N_{tot}$ is the total number of bubbles.}
\end{table}

Definitions and values of various moments of the scaling regime distributions $p(n)$ and $F(n)$ are listed in Table~\ref{bubbleaverages}.  The average side number is $\langle n\rangle = 5.92\pm0.01$, which is slightly less than the value of 6 required by topological reasons for an infinite system.  The area-weighted average side number is somewhat greater, $\langle\langle n \rangle\rangle = 6.53\pm0.08$.  This result is important because, from Eq.~(\ref{avgareaeqn}), the expected average coarsening rate in the scaling regime is proportional to $[\langle\langle n \rangle\rangle -6]$.  The variance of $p(n)$ is $\mu_2=1.56\pm0.02$, consistent with prior scaling-state measurements \cite{StavansGlazier89}.  This quantity is often used as a measure of disorder.  The variance of $F(n)$ is slightly larger, $\nu_2=1.67\pm0.09$.

\begin{figure}
\includegraphics[width = 3.00in]{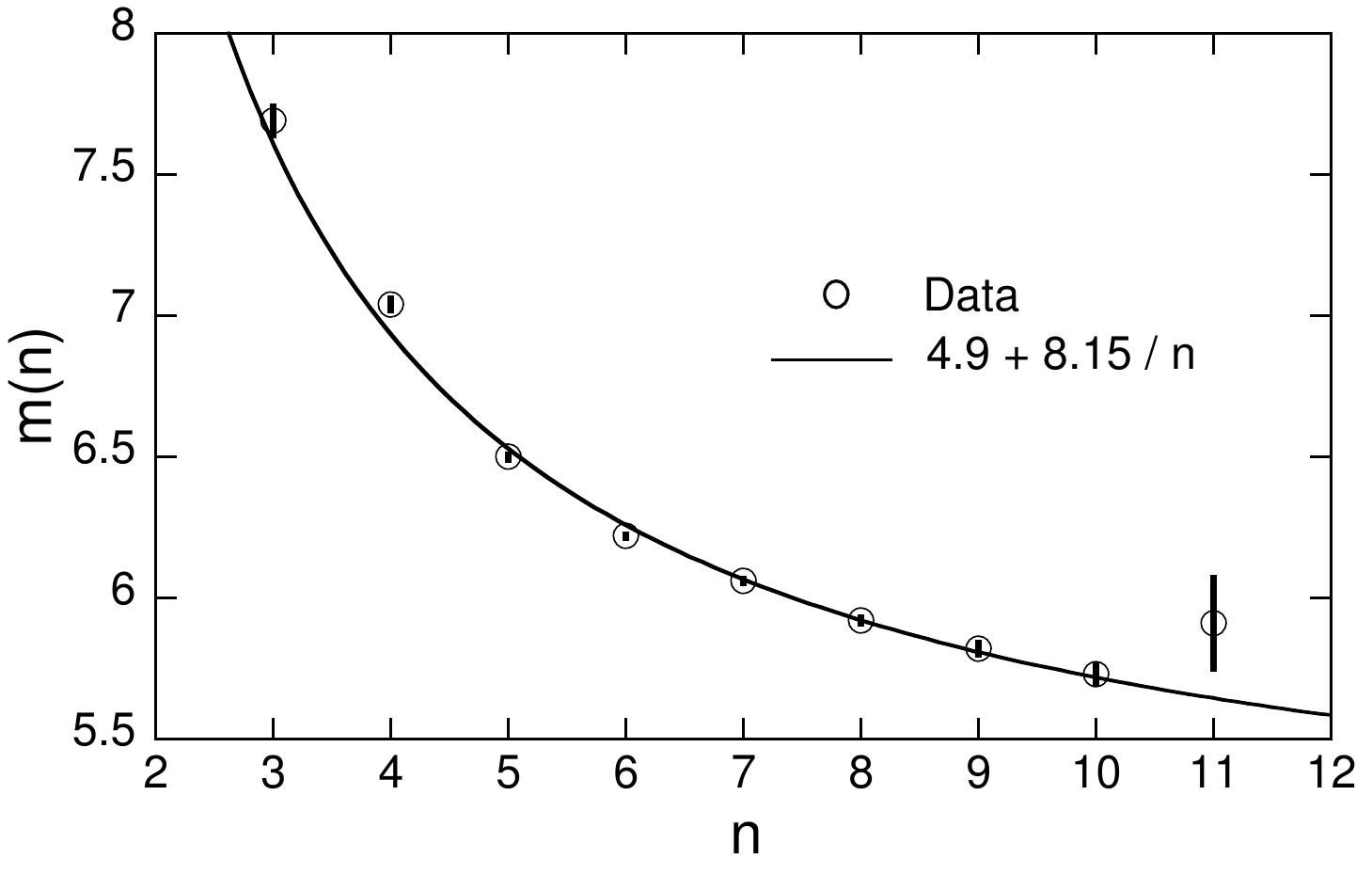}
\caption{The average number $m(n)$ of sides of the neighbors of $n$-sided bubbles.  The results here are an average over all times and over all liquid contents.  The black line is the empirical Aboav-Weaire law, $m(n) = (6-a) + (6a + \mu_{2})/n$, where $\mu_{2}$ is the measure the variance and $a$ is the only fitting parameter, which is found to be $a = 1.1 \pm 0.1$.  The plotted $m(n)$ values are listed in Table~\ref{distributiondata}.}
\label{mn}
\end{figure}

The final purely topological quantity we consider is the average number $m$ of sides of the neighbors of an $n$-sided bubble.  As done for the side distributions, we first verify that $m(n)$ data are independent of time and liquid content and hence may be averaged together.  The final results are displayed in Fig.~\ref{mn}.  For comparison, we obtain a satisfactory fit to the empirical Aboav-Weaire form, $m(n) = (6-a) + (6a + \mu_{2})/n$ \cite{WeaireHutzlerBook}, where $\mu_2=1.56$ is the measured variance and the one fitting parameter is found to be $a=1.1\pm0.1$.   Similar values of $a$ have been found for many cellular patterns \cite{WeaireHutzlerBook}, including two-dimensional foams.

\subsection{Size}

\begin{figure}
\includegraphics[width = 3.00in]{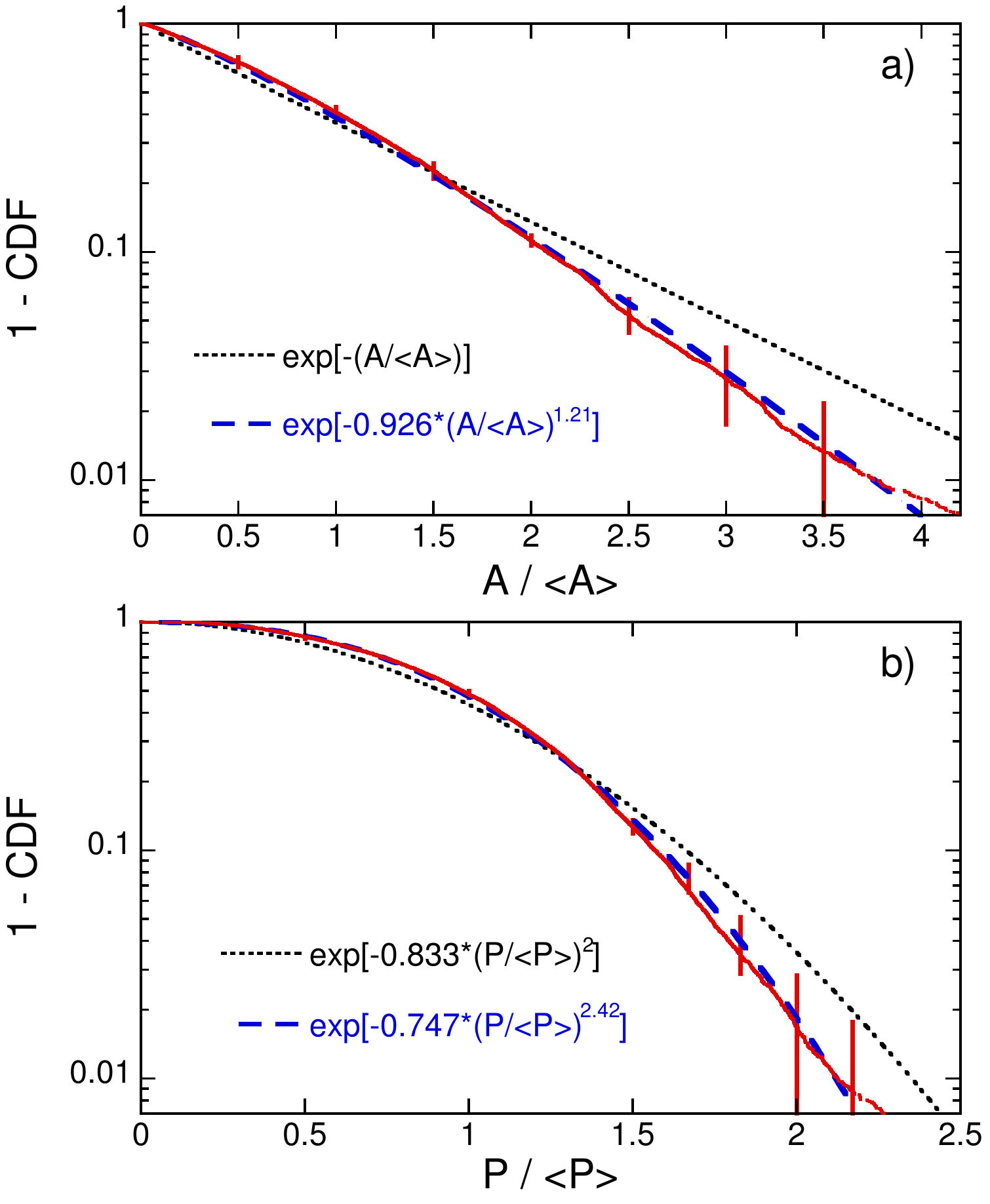}
\caption{(Color online) Cumulative distribution function data for (a) bubble area $A$ and (b) bubble perimeter $P$, averaged over all times and liquid contents.   The error bars represent the uncertainty in the mean, as estimated from the range in values for different liquid contents.  In (a) the black dotted line represents an exponential area distribution and the blue dashed curve represents a compressed exponential.  The corresponding forms for the cumulative perimeter distributions are shown in (b) using the same line codes, further assuming $A\propto P^2$ with the same proportionality constant for all bubbles.}
\label{areaperimdist}
\end{figure}

In this subsection we consider distributions of bubble sizes, beginning with area since this is the quantity that appears in von~Neumann's law.  As a prelude we verify that the distributions are independent of both liquid content and time, when the average is scaled out.  This reinforces the above conclusion that the foam is in a scaling regime, and allows us to combine the time-averaged scaled distributions for each foam sample into a single curve.  The results for one minus the cumulative area distribution are plotted on semi-logarithmic axes in Figs.~\ref{areaperimdist}a.  Error bars are given by the range of values for different liquid contents, divided by the square root of the number of different liquid contents measured.  The data exhibit a slight but nonzero downward curvature, and hence are not quite exponential.  This is consistent with prior work \cite{Flyvbjerg93, Stavans93, Stavans93sf, GlazierWeaire92, GlazierAndersonGrest90}.  A good fit is found to a compressed exponential, given along with the corresponding probability distribution function as
\begin{eqnarray}
{\rm CDF} &=& 1 - e^{ - \left[ \Gamma\left(1+{1\over \alpha}\right) {A \over \langle  A \rangle} \right]^\alpha }    \label{areacdf} \\
{\rm PDF} &=& \alpha\Gamma\left(1+{1\over \alpha}\right)^\alpha\left({A \over \langle  A \rangle}\right)^{\alpha-1} 
                   e^{ - \left[ \Gamma\left(1+{1\over \alpha}\right){A \over \langle  A \rangle} \right]^\alpha }    \label{areapdf}
\end{eqnarray}
with fitting parameter $\alpha=1.21\pm0.05$; this, and a simple exponential (case $a=1$), are both shown in Fig.~\ref{areaperimdist}.

Recall from Eq.~(\ref{avgareaeqn}) that the value of $\langle A \rangle ^{2} / \langle A^{2} \rangle$ helps set the rate of change of the average bubble area in the scaling regime.   Averaging over all times and liquid fractions we find $\langle A \rangle ^{2} / \langle A^{2} \rangle = 0.58 \pm 0.09$, which is close to the value of $1/2$ for a perfectly exponential distribution.  Combining this with the result $\langle\langle n \rangle\rangle = 6.53 \pm 0.08$,  Eq.~(\ref{avgareaeqn}) thus gives the average rate of coarsening for a 2d foam in the self-similar scaling regime as ${\rm d}\langle A\rangle/{\rm d}t = (0.61\pm0.13)K$ where $K$ is the constant in von~Neumann's law for individual bubbles, ${\rm d} A/{\rm d}t = K(n-6)$.  For the vertical cell, the value of $K_0$ then gives the expectation ${\rm d}\langle A\rangle/{\rm d}t = (0.74\pm0.15)$~mm$^2$/hr, which is consistent with the direct measurement of ${\rm d}\langle A\rangle/{\rm d}t = (0.83\pm0.03)$~mm$^2$/hr.

Since bubbles are not all identical in shape, bubble size is not uniquely specified by area.  So next we consider bubble perimeter, which is also important since in two-dimensions coarsening is ultimately driven by a reduction of the total sum of bubble perimeters.  The cumulative distribution for perimeter, averaged over all times and liquid contents, is plotted in Fig.~\ref{areaperimdist}b.  For comparison, we also plot the expectation corresponding to the fitted cumulative area distribution.  For this we must make the further assumption that bubble shape is constant, which implies $A=cP^2$ and $\langle A \rangle = c\langle P^2\rangle$ where $c$ is some constant.  Thus the trial perimeter cumulative distribution function is given by Eq.~(\ref{areacdf}) with $A/\langle A\rangle$ replaced by $P^2/\langle P^2\rangle = [\langle P\rangle^2/\langle P^2\rangle][P/\langle P\rangle]^2$.  From the list of bubble perimeters, we directly compute the second moment to be $\langle P^2 \rangle / \langle P\rangle^2 = 1.20\pm0.06$.  The resulting compressed exponential cumulative perimeter distribution is plotted in Fig.~\ref{areaperimdist}b, and found to agree extremely well with the data.  This foreshadows a point to made directly in a later section: the average bubble shape is remarkably constant.

\subsection{Size-topology}

\begin{figure}
\includegraphics[width = 3.00in]{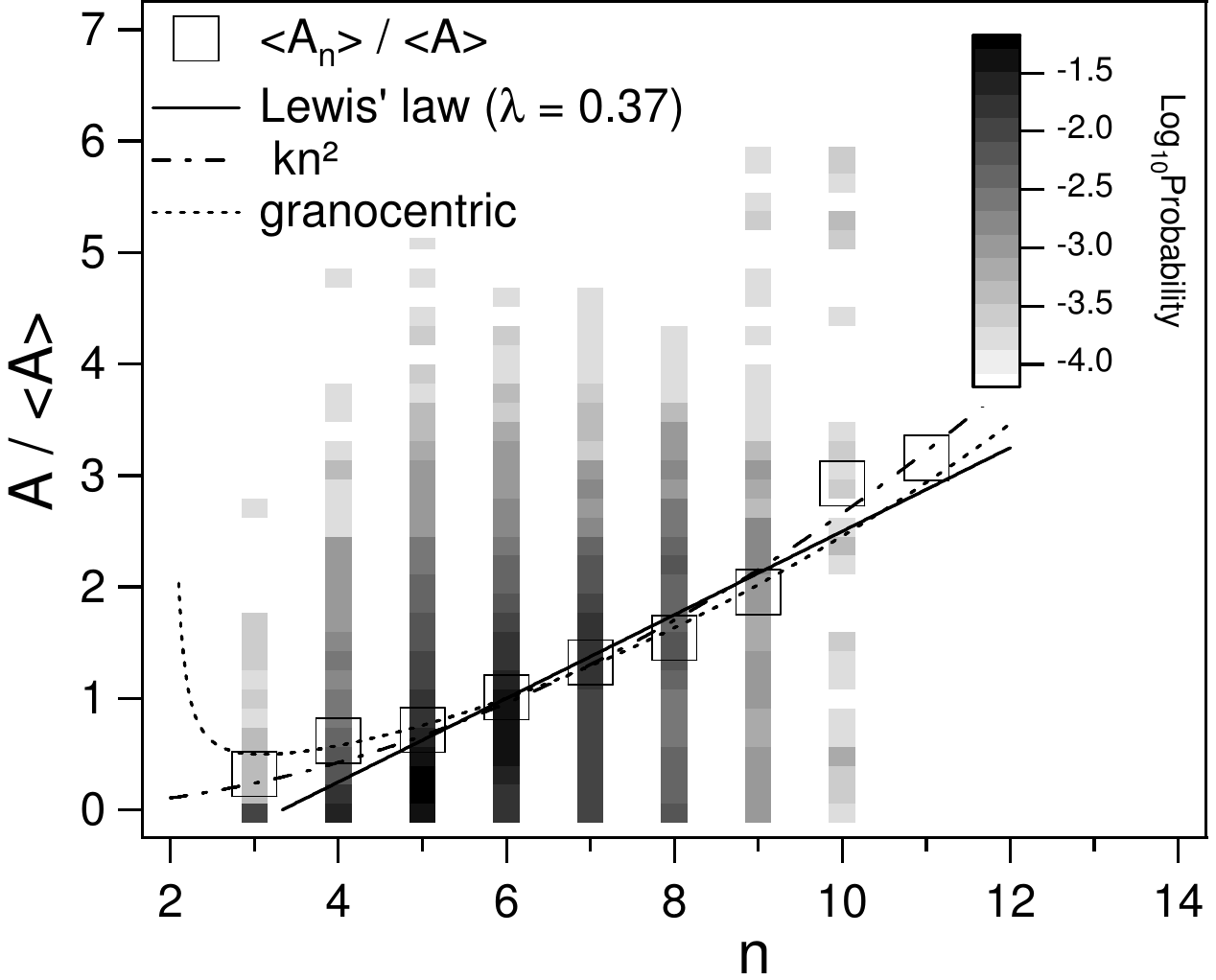}
\caption{Normalized area versus side number for all bubbles; the distribution is given in grayscale and the average is indicated by open squares.  Note that the distributions are quite skewed, as expected since the area distribution averaged over all $n$ is nearly exponential.  The fit to Lewis' law, Eq.~(\protect{\ref{LewisLawEqn}}) with fitting value $\lambda = 0.37 \pm 0.03$, is shown by a solid line.  The fit to the Ref.~\cite{Graneretal11} form, $kn^2$ with fitting value $k = 0.027$, is shown by the dash-dot curve.  The simplified granocentric model prediction, Eq.~(\protect{\ref{granoA}}) with no fitting parameters, is shown by the dotted curve.}
\label{lewislaw}
\end{figure}

\begin{figure}
\includegraphics[width = 3.00in]{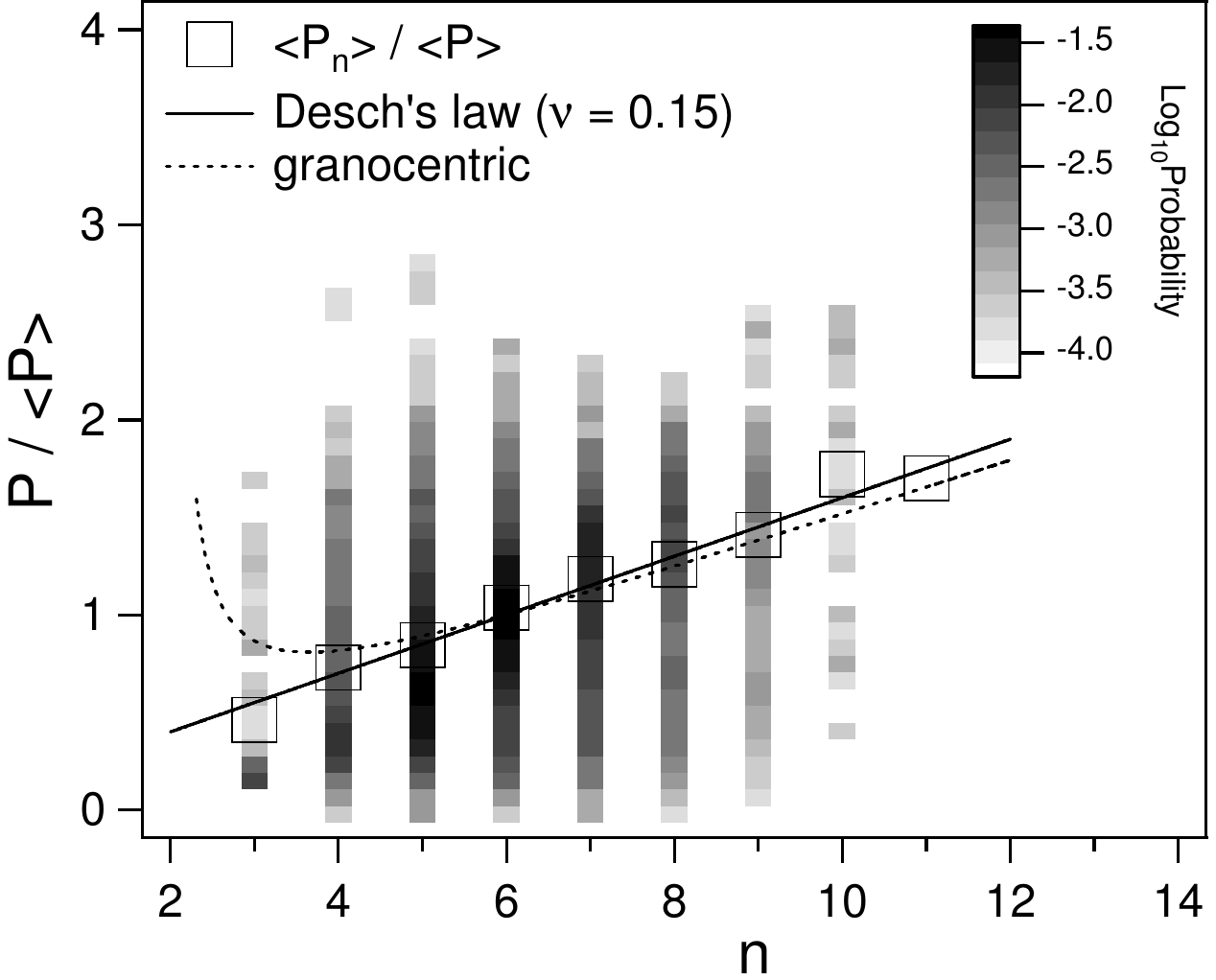}
\caption{Normalized perimeter versus side number for all bubbles; the distribution is given in grayscale and the average is indicated by open squares.  Note that the distributions is quite symmetric around the average.  The fit to Desch's law, Eq.~(\protect{\ref{DeschLawEqn}}) with fitting value $\nu = 0.15 \pm 0.01$, is shown by a solid line.  The simplified granocentric model prediction, Eq.~(\protect{\ref{granoP}}) with no fitting parameters, is shown by the dotted curve.}
\label{deschlaw}
\end{figure}

\begin{table*}[width = 3.00in]
\begin{ruledtabular}
\begin{tabular}{cccccc}
$n$ & $N$ & $\langle A_{n} \rangle / \langle A \rangle \pm \sigma \pm \sigma / \sqrt{N}$ & $\langle P_{n} \rangle / \langle P \rangle$ & $E(n)$ & $C(n)$ \\
\colrule
3 & 195 & $0.32 \pm 0.59 \pm 0.04$ & $0.46 \pm 0.42 \pm 0.03$ & $1.065 \pm 0.02 \pm 0.001$  & $0.19 \pm 0.13 \pm 0.01$ \\
4 & 1217 & $0.62 \pm 0.84 \pm 0.02$ & $0.73 \pm 0.46 \pm 0.01$ & $1.071 \pm 0.02 \pm  0.002$ & $0.299 \pm 0.14 \pm 0.004$ \\
5 & 4462 & $0.72 \pm 0.70 \pm 0.01$ & $0.843 \pm 0.38 \pm 0.006$ & $1.070 \pm 0.03 \pm 0.001$ & $0.137 \pm 0.10 \pm 0.002$  \\
6 & 4634 & $1.01 \pm 0.66 \pm 0.01$ & $1.037 \pm 0.35 \pm 0.005$ & $1.069 \pm 0.03 \pm 0.001$ & $-0.042 \pm 0.18 \pm 0.003$  \\
7 & 2611 & $1.32 \pm 0.77 \pm 0.02$ & $1.183 \pm 0.40 \pm 0.008$ & $1.070 \pm 0.02 \pm 0.001$ & $-0.215 \pm 0.28 \pm 0.005$ \\
8 & 1120 & $1.55 \pm 0.96 \pm 0.03$ & $1.26 \pm 0.47 \pm 0.01$ & $1.070 \pm 0.03 \pm 0.002$ & $-0.409 \pm 0.16 \pm 0.005$ \\
9 & 327 & $1.95 \pm 1.3 \pm 0.07$ & $1.41 \pm 0.54 \pm 0.03$ & $1.067 \pm 0.04 \pm 0.002$ & $-0.59 \pm 0.25 \pm 0.01$ \\
10 & 89 & $2.9 \pm 2.0 \pm 0.2$ & $1.72 \pm 0.70 \pm 0.08$ & $1.065 \pm 0.01 \pm 0.001$ & $-0.73 \pm 0.44 \pm 0.05$ \\
11& 8 & $3.2 \pm 3.7 \pm 1.3$ & $1.7 \pm 1.2 \pm 0.4$ & $1.066 \pm 0.01 \pm 0.005$ & $-0.9 \pm 0.37 \pm 0.1$ \\
\end{tabular}
\end{ruledtabular}
\caption{\label{sizetopo} Shape quantities, averaged over all times and liquid contents, for each side number $n$.  The standard deviation of the distribution ($\sigma$) and the uncertainty in the mean ($\sigma/\sqrt{N}$ where $N$ is the number of bubbles) are also given.  The first two quantities are the area and perimeter, normalized by the average over the whole sample.  The second two quantities are the elongation and circularity, defined by Eqs.~(\protect{\ref{elongdef}-\ref{circdef}}).}
\end{table*}

With topology and size statistics now in hand, we turn to the relationship between these measures.  For many cellular systems, a linear correlation has been observed between either area or perimeter and side number \cite{ChiuReview}:
\begin{eqnarray}
	\langle A_n \rangle / \langle A \rangle &=& 1 + \lambda (n - 6), \ \ \ \  {\rm (Lewis)} \label{LewisLawEqn} \\
	\langle P_n \rangle / \langle P \rangle &=& 1+  \nu (n - 6 ), \ \ \ \  {\rm (Desch)} \label{DeschLawEqn}
\end{eqnarray}
where $\lambda$ and $\nu$ are parameters characteristic to a particular system.  The first of these empirical laws was found by Lewis for epithelial cucumber cells, and is known as Lewis' law \cite{Lewis1928, Lewis1930}.  If $\langle A_{n} \rangle / \langle A \rangle$ is linear in $n$, then it must have this form, but to prove linearity requires additional constraints \cite{RivierLissowski82}.  The analogous relationship for perimeter is called Desch's law or Feltham's law.  If the energy area of a cell is proportional to its perimeter, then entropy is maximized when Desch's law is satisfied \cite{Rivier85}.  Such size-topology relations continue to be a subject of active research \cite{SzetoTam95, Saraivaetal09, Graneretal11, Lambert2012, BrujicPRL12}.

To compare the Lewis and Desch laws with our scaling-state foams, we accumulate time-average statistics for areas and perimeters separately for each side number.  The averages are given in Table~\ref{sizetopo} and are plotted versus $n$ as open squares in Figs.~\ref{lewislaw}-\ref{deschlaw}, respectively.  The scaled average area and perimeter are both indistinguishable from 1 for $n=6$, and both grow with $n$ since larger bubbles tend to have more sides.  For area, the dependence is noticeably faster than linear; for perimeter, the dependence is indistinguishable from linear.  Thus the Desch law provides a better description of scaling regime foams than the Lewis law, as seen by displayed fits.  Indeed the average area data are better fit to $\langle A_n \rangle / \langle A\rangle = (0.027 \pm 0.001)n^2$, in accordance with some simulations and experiments \cite{SzetoTam95, Graneretal11}.  The perimeter data are well fit to the Desch law with $\nu=0.15\pm0.01$.  This is somewhat smaller than previous experimental measurements of $\nu=0.29$ \cite{SzetoTam95}, and $\nu=0.19$ \cite{Graneretal11}.

Apart from the behavior of the averages, the correlation of side number with perimeter has an advantage over area because of the shapes of the distributions, which are also displayed in Figs.~\ref{lewislaw}-\ref{deschlaw} in grayscale.  For area, these are skewed so that the mode is significantly smaller than the average, especially for small $n$ where the peak is near zero as for an exponential distribution.  For perimeter, by contrast, the individual distributions are more symmetrically peaked so that the mode coincides closely with the average.

Regarding the deviation from Lewis' law, it is predicted that this is associated with deviation of the area distribution from exponential \cite{Lambert2012}.  Indeed the area distribution data in Fig.~\ref{areaperimdist}a are not quite exponential.  Further insight into the deviation from the Lewis law has been gained from the granocentric model \cite{BrujicNature09}.  In a simplified version \cite{BrujicPRL12}, a Voronoi-type construction is made for a central particle uniformly surrounded by $n$ equidistant neighbors of the same same size.  This gives the following size-topology relations, without any parameters:
\begin{eqnarray}
	A_n  / \langle A \rangle &=& n / [4\sqrt{3}\sin(2\pi/n)], \label{granoA} \\
	P_n  / \langle P \rangle &=& n / [4\sqrt{3}\cos(\pi/n)]. \label{granoP}
\end{eqnarray}
The first of these is Eq.~(7) from Ref.~\cite{BrujicPRL12} and the second we derived in analogy.  The angle brackets have been removed from $A_n$ and $P_n$ because for a given $n$ there is no distribution in this version of the granocentric model.  These forms are included in Figs.~\ref{lewislaw}-\ref{deschlaw}, and agree quite well with the data.

\subsection{Shape}

\begin{figure}
\includegraphics[width = 3.00in]{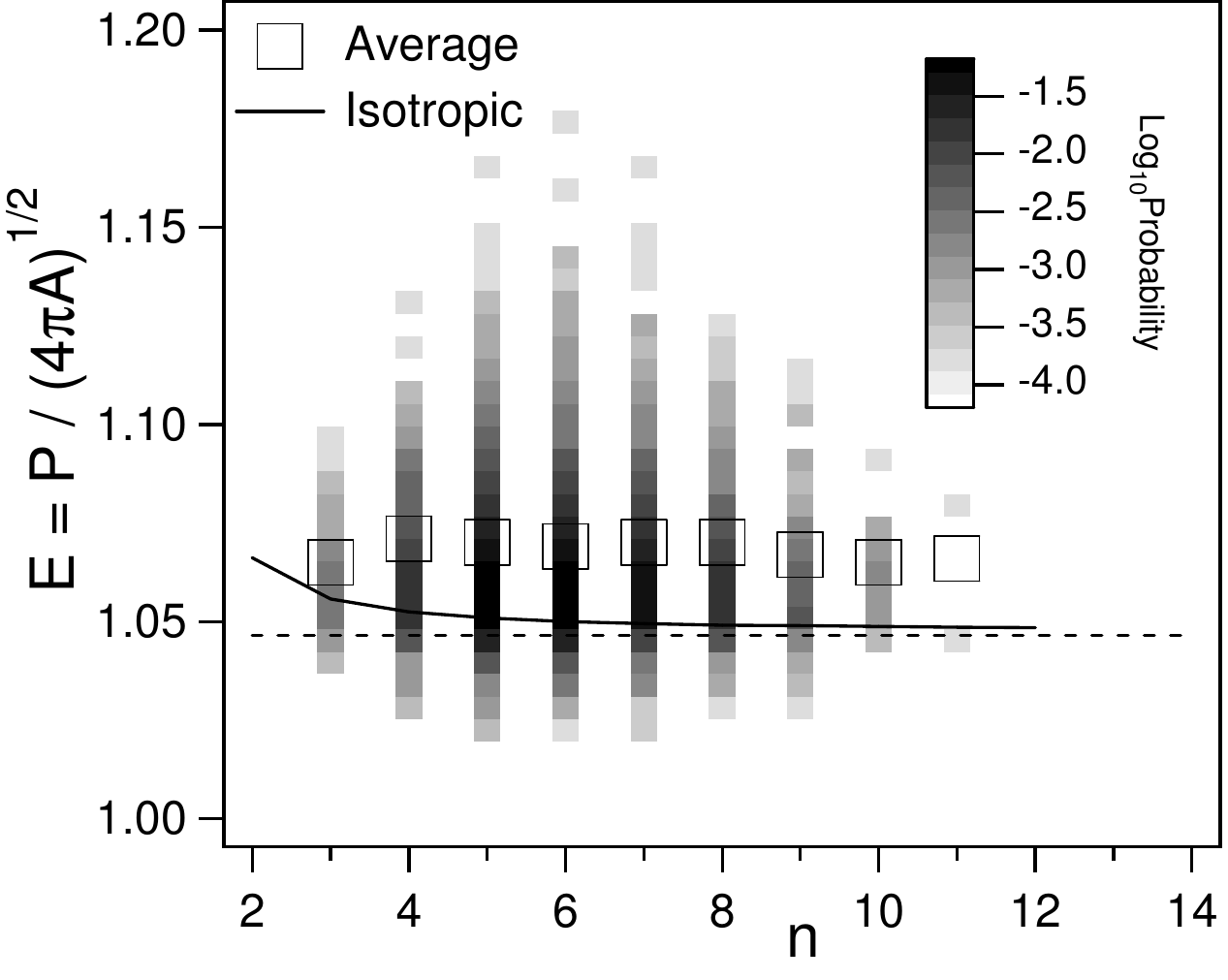}
\caption{Elongation versus side number $n$, averaged over all times and liquid contents, where $P$ is bubble perimeter and $A$ is bubble area.  The probability distribution is shown in grayscale, and the average is shown by open squares.  The solid line represents the elongation for isotropic bubbles, given by  Eq.~(\protect{\ref{isoelong}}); the horizontal dashed line represents $\pi/3=1.047$, the limit as $n\rightarrow\infty$.}
\label{Eshape}
\end{figure}

\begin{figure}
\includegraphics[width = 3.00in]{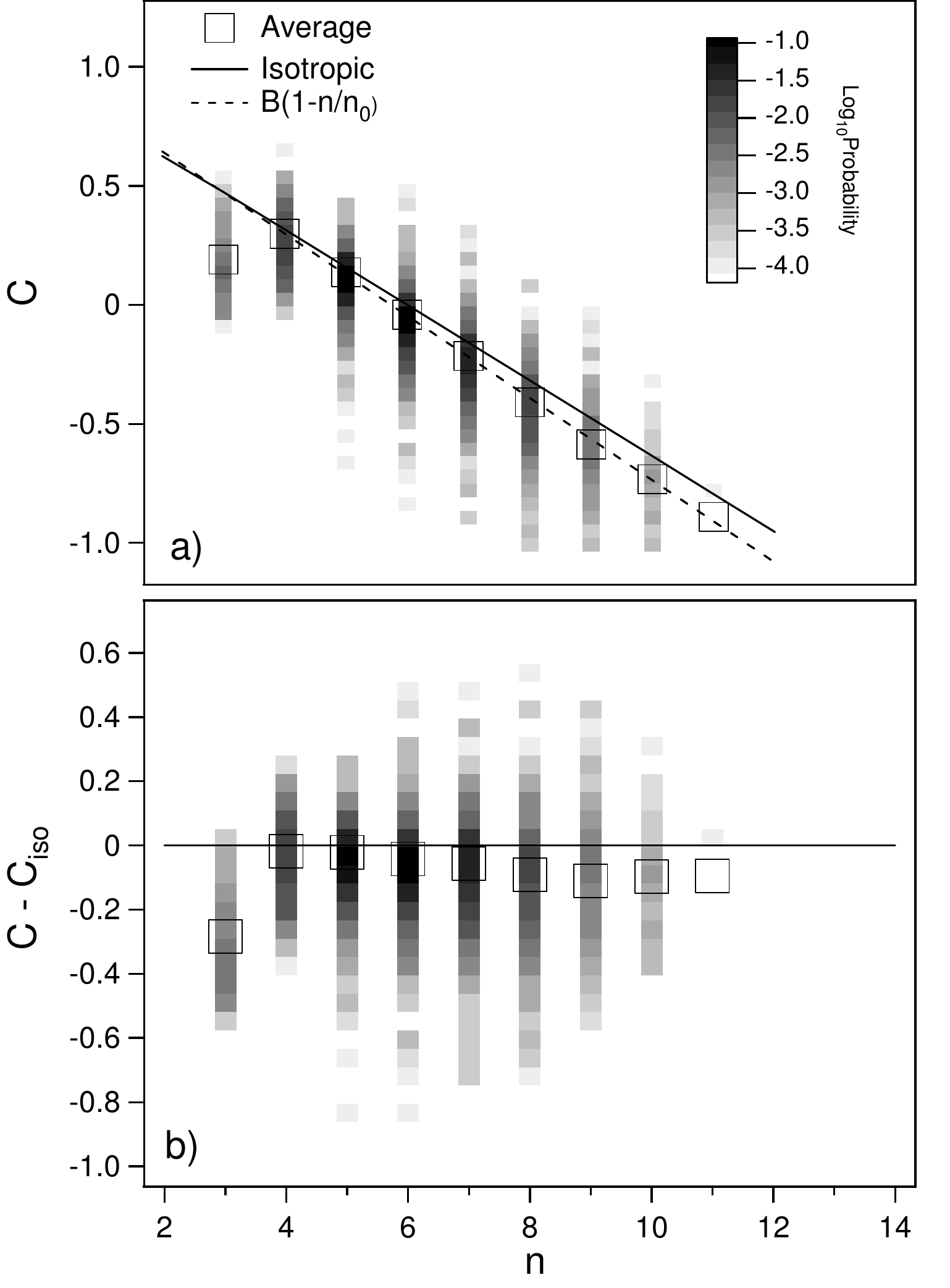}
\caption{(a) Circularity $C$, defined by Eq.~(\protect{\ref{circdef}}), versus side number $n$, averaged over all times and liquid contents.  The probability distribution is shown in grayscale, and the average is shown by open squares.  The solid curve represents the circularity for isotropic bubbles, given by Eq.~(\protect{\ref{isocirc}}).  The dashed line is a fit to $B(1-n/n_0)$, which gives $B=0.99\pm0.02$ and $n_0=5.73\pm0.04$ when the $n=3$ data are excluded.  (b) The difference in circularity between actual and isotropic bubbles.}
\label{Cshape}
\end{figure}

The bubbles in a foam have a wide variety of shapes, even for a given number of sides.  Two shape descriptors that we find in the next section to be relevant for coarsening dynamics are the elongation and circularity, which we define respectively as
\begin{eqnarray}
 E &=& {P / \sqrt{4\pi A}} \label{elongdef}, \\
 C &=& \left( {1\over n}\sum_i^n {1 / R_i} \right) \sqrt{A/\pi}, \label{circdef}
\end{eqnarray}
where $P$ is perimeter, $A$ is area, and $R_i$ is the radius of curvature for the $i^{\rm th}$ side of an $n$-sided bubble.  The sign convention is such that $R_i$ is positive for the bubble on the high-pressure side of the film.  While the quantity $1/E^2$ is commonly known as ``compactness'', we follow Ref.~\cite{Graner2000} in referring to $E$ as ``elongation".  The term in brackets in Eq.~(\ref{circdef}) is an average curvature, with equal weights independent of side length; it is particular to shapes made from circular arc segments, and does not equal $2\pi/P$.  For a circle, these definitions give a minimum elongation of $E=1$ and a maximum circularity of $C=1$.    Note that $C=0$ holds for any shape made of straight line segments.  The simplified granocentric model treats cells as regular $n$-sided polygons, for which the shape descriptors are $E=\sqrt{(n/\pi)\tan(\pi/n)}$ and $C=0$.

For comparison with data, we compute the shape descriptors for ``isotropic'' or ``regular" bubbles consisting of equal arc segments.  These are like regular polygons but with edges replaced by circular arcs, all of radius $R$, that meet at $120^\circ$ as required by Plateau's laws.  Isotropic bubbles have been used to model both two- \cite{Graner2000} and three-dimensional \cite{HilgenfeldtKraynikKoehlerStone00, HilgenfeldtKraynik04, Glicksman05, GlicksmanRiosLewis07} foams.  We find:
\begin{eqnarray}
	P &=& (\pi/ 3) \left| n-6 \right| R \label{isoperim}, \\
	E &=& \sqrt{   {   (\pi/ 3) (n-6)^2  \over 3n\left[\cot(\pi/n)-\sqrt{3}\right] - 2\pi(n-6) }  }, \label{isoelong} \\
	C &=& \pm \sqrt{  {n\over 4\pi}\left[\cot(\pi/n)-\sqrt{3}\right]-{1\over 6}(n-6) }. \label{isocirc}
\end{eqnarray}
The positive root $C>0$ is taken for $n<6$, and the negative root $C<0$ is taken for $n>6$.  Our expression for the elongation is consistent with Eq.~(A3) of Ref.~\cite{Graner2000}, except that our definition includes a factor of $\sqrt{4\pi}$; it approaches $E=\pi/3=1.047$ in the limit $n\rightarrow \infty$.  For $n\ge3$ our expression for the circularity is within 0.5\% of $C=(\pi^2/12)^{1/4}(1-n/6)=0.95(1-n/6)$, the linear expansion around $n=6$.  Both Eqs.~(\ref{isoelong}-\ref{isocirc}) behave badly for $n\le1$, but approach $E=1$ and $C=1$ in the limit $n\rightarrow0$, as expected for a circular bubble with $n=0$ vertices.

We now compute the shape parameters for all the bubbles in all the collected images.  For both, area is taken from the number of enclosed pixels.  For elongation, perimeter is taken from a LabVIEW routine that interpolates the pixellated boundary of the image.  For circularity, the curvature of each segment is taken from the circle defined by the two endpoints and the average of the three middle-most points.  No systematic deviation was ever observed between such arc segments and the pixellated bubble boundaries.  Collecting all results, we find that both $E$ and $C$ are independent of age and liquid content and hence may be combined for better statistics.  The average elongation and the average circularity are plotted versus side number in Figs.~\ref{Eshape}-\ref{Cshape}, respectively.  The probability distributions are also shown in grayscale, and appear to be peaked fairly symmetrically around the average values.  Remarkably, the average elongation appears to be nearly constant and independent of $n$.  Averaging over at all times and all liquid contents and all side numbers gives an average bubble elongation of $\langle E \rangle = 1.0692\pm 0.0005$ and a variance of $e_{2} = 0.004\pm0.001$.  This is about 50\% more elongated from a circle than for isotropic bubbles, Eq.~(\ref{isoelong}).  The data for circularity is nearly linear in $n$ and agrees fairly well with the expectation for isotropic bubbles, Eq.~(\ref{isocirc}), except for three-sided bubbles.  The difference between actual and isotropic bubble circularities is shown for comparison in Fig.~\ref{Cshape}b.  The circularity data are well fit to $C(n)=B(1-n/n_0)$, which gives $B=0.99\pm0.02$ and $n_0=5.73\pm0.04$ when $n=3$ data are excluded.  The average variance of the circularity distributions is $c_2 = 0.08 \pm 0.01$.

\section{Coarsening Dynamics}

\subsection{Data}

\begin{figure*}
\includegraphics[width=6.00in]{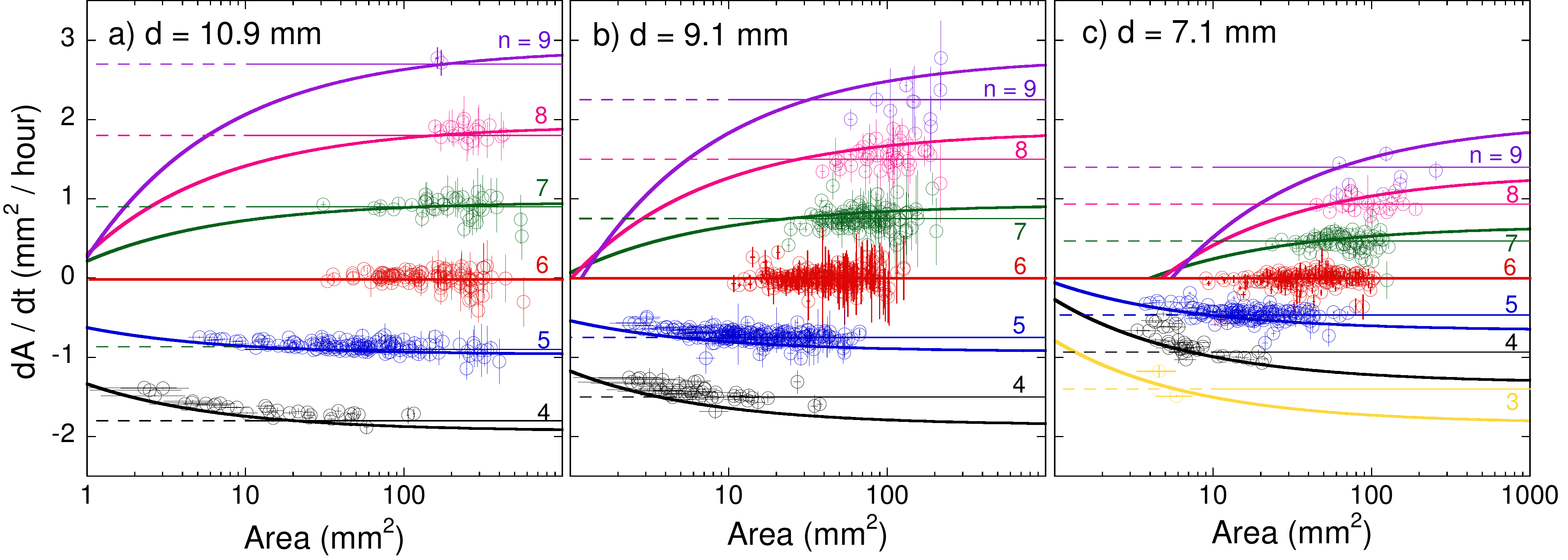}
\caption{(Color online) Rate of area change versus area, for individual bubbles in three foams with increasing liquid content, from left to right, as controlled by the distance $d$ of the foam above the liquid reservoir.  The number $n$ of sides of each bubble is indicated by symbol color, as labeled.  The thin horizontal lines represent a fit to von~Neumann's law, ${\rm d}A/{\rm d}t = K(n-6)$ where $K$ is adjusted fit to the data for $A>10$~mm$^{2}$.  The fitted values of $K$ are plotted versus liquid content Fig.~\ref{kvd}.  The thick curves represent the border-blocking model, Eq.~(\ref{dadt}), where $K_0 = 1.2~{\rm mm}^{2} / {\rm hr}$ is fixed by the data in Fig.~\ref{vertical}, $E$ and $C$ are taken from the averages represented by the open squares in Figs.~\ref{Eshape}-\ref{Cshape},  and $r$ is the only fitting parameter.  The fitted values of $r$ are plotted versus liquid content Fig.~\ref{rvd}.}
\label{coarseningvarea}
\end{figure*}

All measurements discussed so far have been for individual static photographs and have not involved how individual bubbles change over time.  It is also possible to track individual bubbles over time and observe how various quantities change.  This was shown earlier, in Fig.~\ref{avgavt}, for selected bubbles of various $n$ for three liquid contents.   In this plot it can be seen that the rate of change of an $n$-sided bubble's area is slower for wetter foams.

It is possible to measure the area at each time for each bubble in a sequence of images and fit these curves to a line for each bubble.  The slope is ${\rm d}A/{\rm d}t$ for that bubble.  In this way it is possible to measure ${\rm d}A/{\rm d}t$ for a large number of bubbles.  We can then plot ${\rm d}A/{\rm d}t$ against area for a given liquid content.  Examples of this for three different liquid contents are shown in Fig.~\ref{coarseningvarea}.  In these graphs each point is one bubble and the color indicates the number of sides.  The horizontal lines are ${\rm d}A_{n}/{\rm d}t = K (n-6)$ for various $n$ where $K$ is the slope of the proportionality when the data on the plot is plotted as ${\rm d}A/{\rm d}t$ against $n-6$.  On these plots, $K$, the coarsening rate, is the spacing between these horizontal lines.  The values of $K$ are shown against liquid content in Fig.~\ref{kvd}.  The first thing to note is that the coarsening rate decreases as the liquid content increases.  This makes sense as more liquid in the foam should prevent diffusion.  Note also that there is a deviation from von~Neumann's law for small bubbles.  Von~Neumann's law predicts that all bubbles with a given number of sides should coarsen at the same rate; therefore all points of a particular color should fall on the horizontal line of the same color.  Instead, we see that small 4 and 5 sided bubbles fall above the appropriate line, which is to say they are shrinking more slowly than predicted.  Very small bubbles with $n>5$ are not observed because by the time the foam has become two dimensional, there are no very small bubbles with $n>5$ and these bubbles do not shrink, so no examples ever become small enough to observe this effect.  Note also that this deviation appears to be greater for higher liquid contents.  This behavior is explained in the next section.

\subsection{Border-blocking Model}\label{model}

In this section we model the effects of increasing liquid content, both on slowing the coarsening rate and in causing deviation of small bubbles from von~Neumann's law.  To this end we construct a `border blocking' model, with the same assumptions used in the models of Refs.~\cite{BoltonWeaire91, HutzlerWeaire00, StoneKoehlerHilgenfeldt01}.  Namely, the Plateau borders swell with liquid and totally block gas diffusion, reducing the film area and hence slowing the rate of coarsening.  And, as usual, we take the film thickness to be a constant independent of liquid content.  While the prior models dealt only with average growth rates, we now consider the effect of border blocking on individual bubbles through explicit modification of von~Neumann's law.

\begin{figure}
\includegraphics[width=3.00in]{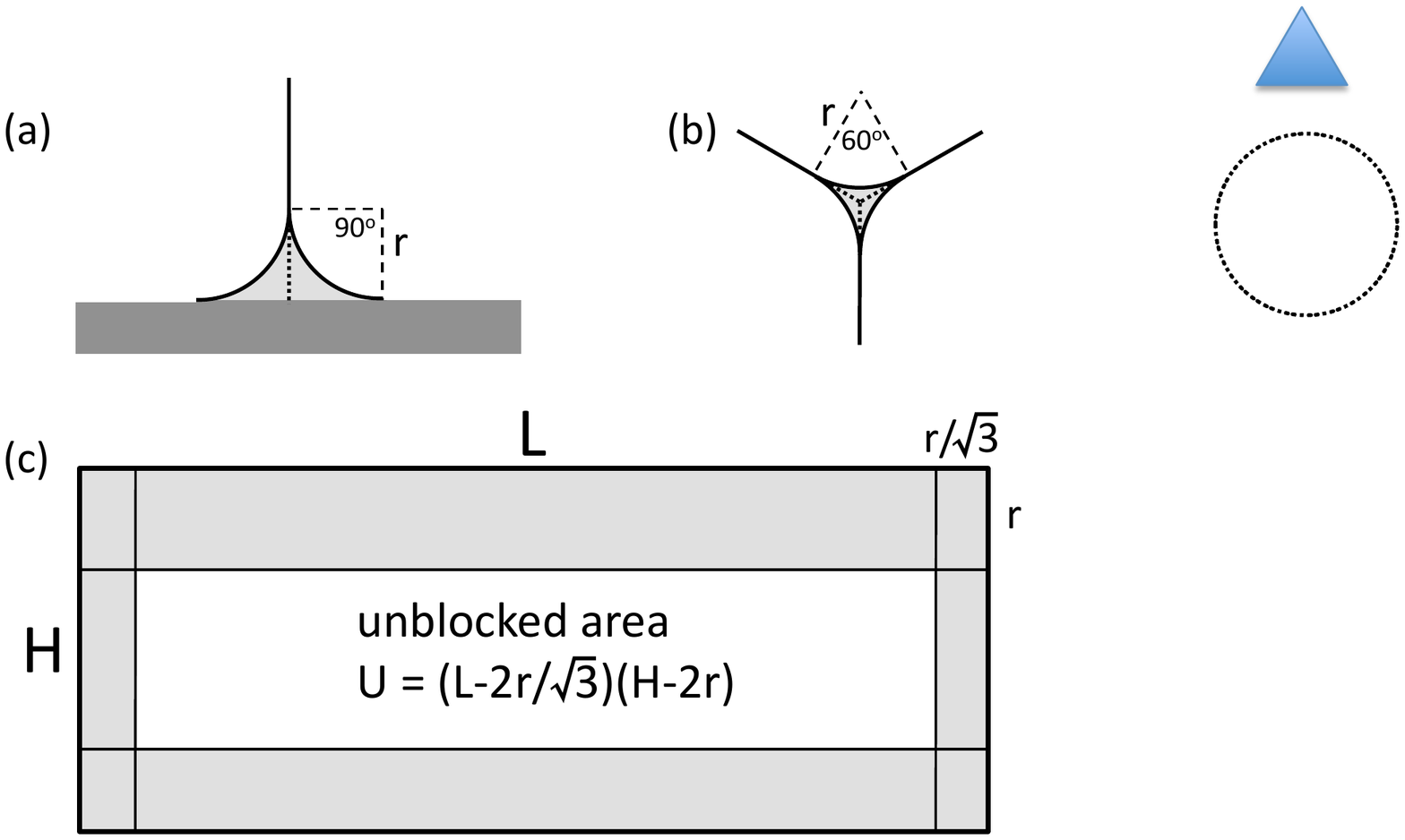}
\caption{Cross section of wet (a) boundary, and (b) vertical, Plateau borders, and also (c) schematic of a vertical soap film.  The Plateau borders have radii of curvature $r$, and are shaded light gray in (a-c).  The blocked portions of the films, though which gas is assumed not to diffuse, are represented by dotted lines, and are seen to have respective length of (a) $r$ and (b) $r\tan(30^\circ)=r/\sqrt{3}$.  As shown in (c), this give the central unblocked area of the film as $U=(L-2r/\sqrt{3})(H-2r)$, where $L$ and $H$ are respectively the length and height of the film in the dry limit $r\rightarrow0$.}
\label{FilmSchematic}
\end{figure}

The rate ${\rm d}V/{\rm d}t$ at which a bubble's volume changes with time is proportional to the sum of the gas diffusion rates across all its films.  And the gas diffusion rate across each film is proportional to the Laplace pressure difference and the film area.  For the quasi-2d experiments here, vertical soap films span the gap $H$ between plates and have constant radius of curvature $R$ along the plates.  As the starting point, we therefore take
\begin{equation}
	{\rm d}V/{\rm d}t \propto -\sum_i(\gamma/R_i)U_i,
\label{dvdtmodel}
\end{equation}
where the Laplace pressure $\gamma/R_i$ is positive for concave films and $U_i$ is the unblocked area through which gas is free to diffuse.  To aid in computing the left and right-hand sides of this expression, we show the salient geometrical features of the Plateau borders and films in Fig.~\ref{FilmSchematic}.  As before, $H$ is the gap between the plates.  And we define $L_i$ as the arclength of the films in the dry limit.  For simplicity we take the radius of curvature of the Plateau borders as $r= \gamma/(\rho g d)$, Eq.~(\ref{rvdequation}), to be the same everywhere -- for the boundary borders at the top and bottom plates and along the vertical borders where three films meet.  We also assume that the vertical Plateau borders are symmetric.  By the decoration theorem, the swelling of Plateau borders with liquid does not affect the soap films -- liquid is merely painted onto the Plateau borders.  Hence $R_i$ and $L_i$ are independent of liquid content.

The first task is to compute the left-hand side of Eq.~(\ref{dvdtmodel}), ${\rm d}V/{\rm d}t$, in terms of the observable skeletonized bubble area $A$.  From the schematic diagram in Fig.~\ref{FilmSchematic}a, it may be seen that the boundary Plateau borders have cross sectional area $(1-\pi/4)r^2$ inside each bubble.  Thus the bubble volume is $V=AH - (1-\pi/4)r^2\cdot(2P)$, minus smaller terms due to vertical Plateau borders and vertices.  And the bubble perimeter may be expressed from the definition of elongation as $P=\sqrt{4\pi A}E$.  All this gives
\begin{equation}
	{{\rm d}V \over {\rm d}t } = H {{\rm d}A \over {\rm d}t}\left[ 1 - \left(1-{\pi\over 4}\right){ \sqrt{4\pi}Er^2 \over H\sqrt{A} } \right].
\label{dvdt}
\end{equation}
For wet foams, bubble volume is not quite proportional to bubble area; the correction depends on shape and is more important for wetter foams and smaller bubbles.

To compute the unblocked film area $U$ as a function of liquid content, note from from Fig.~\ref{FilmSchematic}a that the length of film blocked by a boundary border is simply $r$.  And from Fig.~\ref{FilmSchematic}b the length of film blocked by vertical border is $r \tan(30^\circ)=r/\sqrt{3}$.  Each film is thus blocked by $r$ along top and bottom and by $r/\sqrt{3}$ along the sides, as shown in Fig.~\ref{FilmSchematic}c.  Thus the unblocked area is $U=(H-2r)(L-2r/\sqrt{3})$, where $L$ is the arc length of the curved film measured along the plates between centers of the swollen vertical borders (i.e. the films length as measured in the dry limit).  The right-hand side of Eq.~(\ref{dvdtmodel}) is thus
\begin{eqnarray}
\sum_i {\gamma\over R_i} U_i &\propto& - \sum_i {\gamma\over R_i} (H-2r)(L_i-{2r / \sqrt{3}} ), \\
 &\propto& -\left(1-{2r\over H}\right) \sum_i \left( {L_i\over R_i} - {2r\over \sqrt{3}R_i} \right).
\label{dadtprop}
\end{eqnarray}
As in the usual derivation of von~Neumann's law, the sum of turning angles around a bubble is $2\pi = \sum_i [ (L_i/R_i)+\pi/3]$, since films in the dry limit are circular arcs that subtend angle $L_i/R_i$ and meet at angles of $2\pi/3$ at the center of the inflated vertical Plateau borders.  The latter follows from the decoration theorem, which holds since vertical Plateau borders do not merge \cite{BoltonWeaire91}.  Therefore the first quantity being summed in Eq.~(\ref{dadtprop}) is $\sum_i(L_i/R_i) = (\pi/3)(6-n)$.  The other quantity being summed may be expressed as $\sum(1/R_i)=nC/\sqrt{A/\pi}$ by the definition of circularity.

Combining all the above ingredients we arrive at the final prediction for the rate of area change:
\begin{equation}
\frac{{\rm d}A}{{\rm d}t}  = K_0 {    \left(1-\frac{2r}{H}\right)\left[\left(n-6\right) + {6Cn r \over \sqrt{3\pi A}}\right]  \over  
                                                             1-\left(1-{\pi\over 4}\right){ \sqrt{4\pi}E r^2 \over H\sqrt{A} }            }
\label{dadt}
\end{equation}
where $K_0$ is the proportionality constant in von~Neumann's law for a perfectly dry foam with $r=0$ (see Eq.~(\ref{VN})).  Note that the overall coarsening rate is reduced with liquid content by a factor $(1-2r/H)$ that is the same for all bubbles.  However there are also two terms that depend on the {\it shape} of the bubble, via circularity $C$ and elongation $E$, and that cause deviation from the usual $(n-6)$ von~Neumann behavior.  Both of these terms become more important for wetter foams and smaller bubbles.

Before comparing Eq.~(\ref{dadt}) with data, we first emphasize the assumptions on which it is based.  First, it incorrectly assumes that the liquid in the Plateau borders totally blocks the diffusion of gas; rather, gas can diffuse through borders too, but at a slower rate.  Second, it assumes that the liquid in the vertical Plateau borders does not cause deviation in the angles from Plateau's laws, i.e. that the decoration theorem holds.  This should be valid, as discussed above.  Third, for simplicity, it incorrectly assumes that radius $r$ of the borders is constant; rather, it decreases continuously as a function of the height above the liquid reservoir.  Despite these issues, we show next that the model fits the data well and explains the deviation from von Neumann's law for small wet bubbles.

\subsection{Comparison}

\begin{figure}
\includegraphics[width=3.00in]{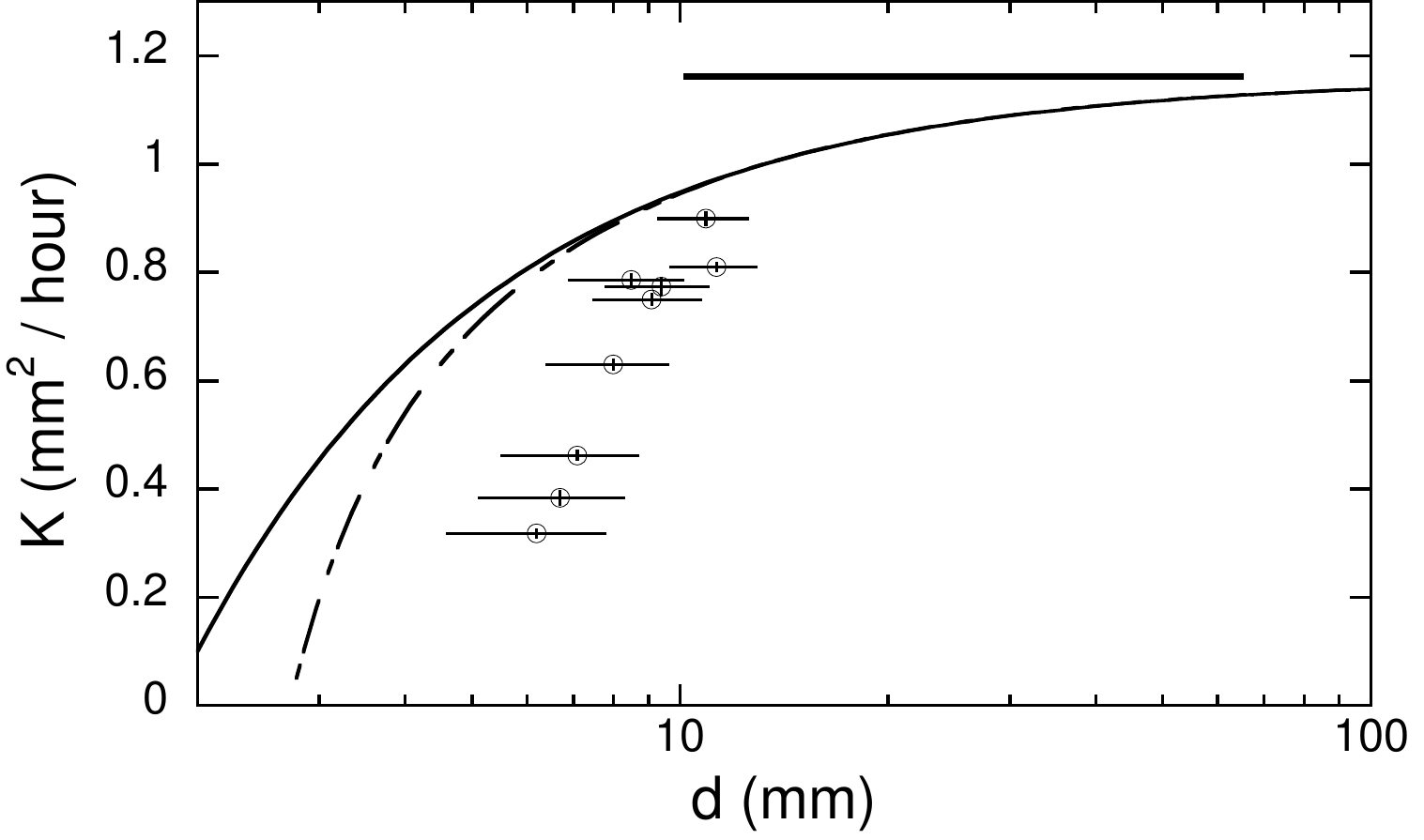}
\caption{Coarsening rate $K = ({\rm d}A_{n}/{\rm d}t) / (n - 6)$, versus height $d$ of the foam above the liquid reservoir.  Values correspond to the thin-line fits in Fig.~\ref{coarseningvarea}.  The solid curve is the predicted relationship $K = K_0 (1 - 2r/H)$.  $K_0$ is the observed coarsening rate for very dry bubbles, $1<d<6~{\rm cm}$, shown as a horizontal line.  This value corresponds to the fit in Fig.~\ref{vertical}.   The dashed curve is the expected average $K$ if the top and bottom plates have different $r$, owing to the gap $H$ of the sample cell.}
\label{kvd}
\end{figure}

\begin{figure}
\includegraphics[width=3.00in]{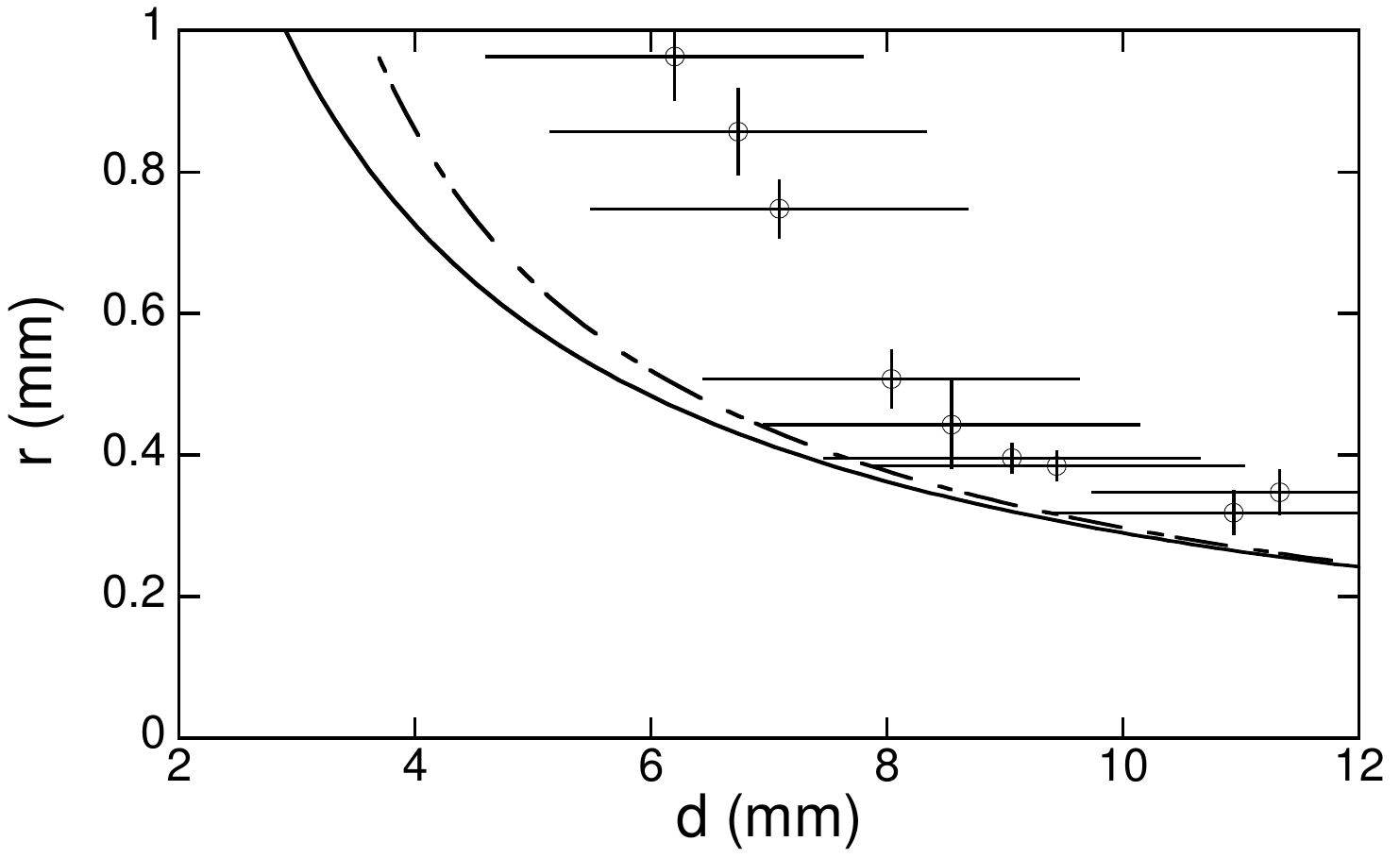}
\caption{Fitted value $r$ versus height of the foam above the reservoir.  Values correspond to the heavy-curve fits in Fig.~\ref{coarseningvarea}.  Solid line is the expected relationship $r = \gamma/\rho g d$.  The dashed line is the expected average $r$ if the top and bottom plates have different $r$, owing to the gap $H$ of the sample cell.}
\label{rvd}
\end{figure}

We analyze our coarsening rate data in two ways.  The first is a standard von~Neumann-type analysis for bubbles large enough that Eq.~(\ref{dadt}) reduces to ${\rm d}A/{\rm d}t = K_0(1-2r/H)(n-6)$, i.e.\ that ${\rm d}A/{\rm d}t=K(n-6)$ holds and is independent of $A$.  For this we plot ${\rm d}A/{\rm d}t$ versus $(n-6)$ for each bubble for a given liquid content, and fit for an overall coarsening rate, $K$.  These fits correspond to the horizontal lines in Fig.~\ref{coarseningvarea}, which show satisfactory von~Neumann behavior for bubbles with area $A > 10~{\rm mm^{2}}$.  The fitting results for $K$ are plotted in Fig.~\ref{kvd} versus the height $d$ of the foam above the liquid reservoir.  The expectation, $K = K_0(1 - 2r/H)$ with $r = \gamma / \rho g d$ and $\gamma = 25~{\rm dynes} / {\rm cm}$, is also shown for comparison.  The trend is correct, but not quantitatively so.  Allowing for $r$ to be different at the top and bottom plates due to their difference in height improves the agreement, which is shown as a dotted line on the graph.

The second analysis is to fit the ${\rm d}A/{\rm d}t$ vs $A$ data shown in Fig.~\ref{coarseningvarea} to the border-blocking prediction, Eq.~(\ref{dadt}), by adjusting only the value of $r$.  The value of $K_0$ is fixed to 1.2~mm$^2$/hr, as found from Fig.~\ref{vertical} for the dry foam limit.   For each $n$, the values of $E$ and $C$ are taken from average elongation and circularity given by the open squares observed in Figs.~\ref{Eshape}-\ref{Cshape}, respectively.  The gap $H$ between the plates is large enough, however, that the term involving $E$ ranges from 0.01 to 0.15 and hence is relatively minor.   Only data for bubbles with $n\le5$ was used to calculate a fit for $r$ because only these bubbles included small bubbles that deviated from von~Neumann's law.  This gives fits such as shown by the heavy curves in Fig.~\ref{coarseningvarea}.  We see that the model fits the coarsening rate data quite well, accurately capturing the deviation from von Neumann's law with a single fitting parameter, $r$.   The fitted values of this parameter are plotted in Fig.~\ref{rvd} versus liquid content and compared with the expectation $r = \gamma / \rho g d$.   The trend and order of magnitude is correct, but the agreement is not very good.  Considering the variation in $r$ due to the height of the cell improves the comparison, but does not seem to account for the full discrepancy.

\section{Conclusion}

In this paper we have presented several advances.  First we devised a novel sample cell that allows the liquid content of Plateau borders to be controlled while maintaining a two-dimensional structure consistent with the decoration theorem.  With this apparatus and digital video imaging, we collected extensive data for bubble statistics and coarsening rates.  Besides the usual side-number and area distributions, we also analyzed for correlations between size and topology and compared with several predictions.  In addition we introduced several new quantities and demonstrated how they are important for the theory of coarsening.  This includes the area-weighted side-number distribution, $F(n)$, and the area-weighted average side number, $\langle\langle n\rangle \rangle$, which have general importance via Eq.~(\ref{avgareaeqn}) for the rate of change of average bubble area in the scaling regime.  This also includes two dimensionless parameters for specifying the shapes of bubbles -- the elongation $E$ and the circularity $C$.  We acquired extensive data on all four of these quantities, of which we are aware of no precedent.  We also acquired extensive data for the rate of coarsening of bubbles, as a function of both side number and -- more novelly -- of liquid content.  We find that increasing wetness causes a deviation from von~Neumann's law, which becomes more important for smaller bubbles.  This behavior we were able to model successfully in terms of an explicit modification of von~Neumann's law to include the blockage of gas diffusion by Plateau border.  An interesting feature of this model is that the bubble shape parameters $E$ and $C$ both appear.  Of the endless ways to quantify shape, these two actually have physical significance for the behavior of the bubbles in foam.  Altogether our work significantly extends the description of the scaling regime of two-dimensional foams, and of the influence of wetness on coarsening.  We hope this might help point the way for future studies of bubble-scale behavior in the coarsening of wet three-dimensional foams.

This work was supported by NASA Microgravity Fluid Physics Grant NNX07AP20G.  We thank J. Bruji\'c and K. Newhall for helpful conversations regarding the granocentric model \cite{BrujicPRL12} and size-topology correlations.

\bibliography{CoarseningRefs}

\begin{thebibliography}{61}
\expandafter\ifx\csname natexlab\endcsname\relax\def\natexlab#1{#1}\fi
\expandafter\ifx\csname bibnamefont\endcsname\relax
  \def\bibnamefont#1{#1}\fi
\expandafter\ifx\csname bibfnamefont\endcsname\relax
  \def\bibfnamefont#1{#1}\fi
\expandafter\ifx\csname citenamefont\endcsname\relax
  \def\citenamefont#1{#1}\fi
\expandafter\ifx\csname url\endcsname\relax
  \def\url#1{\texttt{#1}}\fi
\expandafter\ifx\csname urlprefix\endcsname\relax\def\urlprefix{URL }\fi
\providecommand{\bibinfo}[2]{#2}
\providecommand{\eprint}[2][]{\url{#2}}

\bibitem[{\citenamefont{Weaire and Hutzler}(1999)}]{WeaireHutzlerBook}
\bibinfo{author}{\bibfnamefont{D.}~\bibnamefont{Weaire}} \bibnamefont{and}
  \bibinfo{author}{\bibfnamefont{S.}~\bibnamefont{Hutzler}},
  \emph{\bibinfo{title}{The Physics of Foams}} (\bibinfo{publisher}{Oxford
  University Press}, \bibinfo{address}{New York, NY}, \bibinfo{year}{1999}).

\bibitem[{\citenamefont{Glazier and Weaire}(1992)}]{GlazierWeaire92}
\bibinfo{author}{\bibfnamefont{J.}~\bibnamefont{Glazier}} \bibnamefont{and}
  \bibinfo{author}{\bibfnamefont{D.}~\bibnamefont{Weaire}},
  \bibinfo{journal}{J. Phys.: Condens. Matter} \textbf{\bibinfo{volume}{4}},
  \bibinfo{pages}{1867} (\bibinfo{year}{1992}).

\bibitem[{\citenamefont{Stavans}(1993{\natexlab{a}})}]{Stavans93}
\bibinfo{author}{\bibfnamefont{J.}~\bibnamefont{Stavans}},
  \bibinfo{journal}{Rep. Prog. Phys.} \textbf{\bibinfo{volume}{56}},
  \bibinfo{pages}{733} (\bibinfo{year}{1993}{\natexlab{a}}).

\bibitem[{\citenamefont{von Neumann}(1952)}]{VonNeumann}
\bibinfo{author}{\bibfnamefont{J.}~\bibnamefont{von Neumann}}, in
  \emph{\bibinfo{booktitle}{Metal Interfaces}} (\bibinfo{publisher}{American
  Society for Metals}, \bibinfo{address}{Cleveland}, \bibinfo{year}{1952}), pp.
  \bibinfo{pages}{108--110}.

\bibitem[{\citenamefont{Glazier et~al.}(1987)\citenamefont{Glazier, Gross, and
  Stavans}}]{GlazierGrossStavans87}
\bibinfo{author}{\bibfnamefont{J.~A.} \bibnamefont{Glazier}},
  \bibinfo{author}{\bibfnamefont{S.~P.} \bibnamefont{Gross}}, \bibnamefont{and}
  \bibinfo{author}{\bibfnamefont{J.}~\bibnamefont{Stavans}},
  \bibinfo{journal}{Phys. Rev. A} \textbf{\bibinfo{volume}{36}},
  \bibinfo{pages}{306} (\bibinfo{year}{1987}).

\bibitem[{\citenamefont{Glazier and Stavans}(1989)}]{GlazierStavans89}
\bibinfo{author}{\bibfnamefont{J.~A.} \bibnamefont{Glazier}} \bibnamefont{and}
  \bibinfo{author}{\bibfnamefont{J.}~\bibnamefont{Stavans}},
  \bibinfo{journal}{Phys. Rev. A} \textbf{\bibinfo{volume}{40}},
  \bibinfo{pages}{7398} (\bibinfo{year}{1989}).

\bibitem[{\citenamefont{Stavans and Glazier}(1989)}]{StavansGlazier89}
\bibinfo{author}{\bibfnamefont{J.}~\bibnamefont{Stavans}} \bibnamefont{and}
  \bibinfo{author}{\bibfnamefont{J.~A.} \bibnamefont{Glazier}},
  \bibinfo{journal}{Phys. Rev. Lett.} \textbf{\bibinfo{volume}{62}},
  \bibinfo{pages}{1318} (\bibinfo{year}{1989}).

\bibitem[{\citenamefont{Stavans}(1990)}]{Stavans90}
\bibinfo{author}{\bibfnamefont{J.}~\bibnamefont{Stavans}},
  \bibinfo{journal}{Phys. Rev. A} \textbf{\bibinfo{volume}{42}},
  \bibinfo{pages}{5049} (\bibinfo{year}{1990}).

\bibitem[{\citenamefont{Stavans}(1993{\natexlab{b}})}]{Stavans93sf}
\bibinfo{author}{\bibfnamefont{J.}~\bibnamefont{Stavans}},
  \bibinfo{journal}{Physica A} \textbf{\bibinfo{volume}{194}},
  \bibinfo{pages}{307} (\bibinfo{year}{1993}{\natexlab{b}}).

\bibitem[{\citenamefont{de~Icaza et~al.}(1994)\citenamefont{de~Icaza,
  Jim$\acute{{\rm e}}$nez-Ceniceros, and Casta$\tilde{{\rm n}}$o}}]{Icaza94}
\bibinfo{author}{\bibfnamefont{M.}~\bibnamefont{de~Icaza}},
  \bibinfo{author}{\bibfnamefont{A.}~\bibnamefont{Jim$\acute{{\rm
  e}}$nez-Ceniceros}}, \bibnamefont{and} \bibinfo{author}{\bibfnamefont{V.~M.}
  \bibnamefont{Casta$\tilde{{\rm n}}$o}}, \bibinfo{journal}{J. Appl. Phys.}
  \textbf{\bibinfo{volume}{76}}, \bibinfo{pages}{7317} (\bibinfo{year}{1994}).

\bibitem[{\citenamefont{Krichevsky and Stavans}(1992)}]{StavansKrichevsky92}
\bibinfo{author}{\bibfnamefont{O.}~\bibnamefont{Krichevsky}} \bibnamefont{and}
  \bibinfo{author}{\bibfnamefont{J.}~\bibnamefont{Stavans}},
  \bibinfo{journal}{Phys. Rev. B} \textbf{\bibinfo{volume}{46}},
  \bibinfo{pages}{10579} (\bibinfo{year}{1992}).

\bibitem[{\citenamefont{Rosa and Fortes}(1999)}]{RosaFortes99}
\bibinfo{author}{\bibfnamefont{M.~E.} \bibnamefont{Rosa}} \bibnamefont{and}
  \bibinfo{author}{\bibfnamefont{M.~A.} \bibnamefont{Fortes}},
  \bibinfo{journal}{Philos. Mag. A} \textbf{\bibinfo{volume}{79}},
  \bibinfo{pages}{1871} (\bibinfo{year}{1999}).

\bibitem[{\citenamefont{Rosa et~al.}(2002)\citenamefont{Rosa, Afonso, and
  Fortes}}]{RosaFortes02}
\bibinfo{author}{\bibfnamefont{A.~E.} \bibnamefont{Rosa}},
  \bibinfo{author}{\bibfnamefont{L.}~\bibnamefont{Afonso}}, \bibnamefont{and}
  \bibinfo{author}{\bibfnamefont{M.~A.} \bibnamefont{Fortes}},
  \bibinfo{journal}{Philos. Mag. A} \textbf{\bibinfo{volume}{82}},
  \bibinfo{pages}{2953} (\bibinfo{year}{2002}).

\bibitem[{\citenamefont{Stine et~al.}(1990)\citenamefont{Stine, Rauseo, Moore,
  Wise, and Knobler}}]{KnoblerPRA90}
\bibinfo{author}{\bibfnamefont{K.~J.} \bibnamefont{Stine}},
  \bibinfo{author}{\bibfnamefont{S.~A.} \bibnamefont{Rauseo}},
  \bibinfo{author}{\bibfnamefont{B.~G.} \bibnamefont{Moore}},
  \bibinfo{author}{\bibfnamefont{J.~A.} \bibnamefont{Wise}}, \bibnamefont{and}
  \bibinfo{author}{\bibfnamefont{C.~M.} \bibnamefont{Knobler}},
  \bibinfo{journal}{Phys. Rev. A} \textbf{\bibinfo{volume}{41}},
  \bibinfo{pages}{6884} (\bibinfo{year}{1990}).

\bibitem[{\citenamefont{Berge et~al.}(1990)\citenamefont{Berge, Simon, and
  Libchaber}}]{Bergeetal90}
\bibinfo{author}{\bibfnamefont{B.}~\bibnamefont{Berge}},
  \bibinfo{author}{\bibfnamefont{A.~J.} \bibnamefont{Simon}}, \bibnamefont{and}
  \bibinfo{author}{\bibfnamefont{A.}~\bibnamefont{Libchaber}},
  \bibinfo{journal}{Phys. Rev. A} \textbf{\bibinfo{volume}{41}},
  \bibinfo{pages}{6893} (\bibinfo{year}{1990}).

\bibitem[{\citenamefont{Kermode and Weaire}(1990)}]{KermodeWeaire90}
\bibinfo{author}{\bibfnamefont{J.~P.} \bibnamefont{Kermode}} \bibnamefont{and}
  \bibinfo{author}{\bibfnamefont{D.}~\bibnamefont{Weaire}},
  \bibinfo{journal}{Computer Phys. Comm.} \textbf{\bibinfo{volume}{60}},
  \bibinfo{pages}{75} (\bibinfo{year}{1990}).

\bibitem[{\citenamefont{Glazier et~al.}(1990)\citenamefont{Glazier, Anderson,
  and Grest}}]{GlazierAndersonGrest90}
\bibinfo{author}{\bibfnamefont{J.}~\bibnamefont{Glazier}},
  \bibinfo{author}{\bibfnamefont{M.}~\bibnamefont{Anderson}}, \bibnamefont{and}
  \bibinfo{author}{\bibfnamefont{G.}~\bibnamefont{Grest}},
  \bibinfo{journal}{Philos. Mag. B} \textbf{\bibinfo{volume}{62}},
  \bibinfo{pages}{615} (\bibinfo{year}{1990}).

\bibitem[{\citenamefont{Herdtle and Aref}(1992)}]{HerdtleAref92}
\bibinfo{author}{\bibfnamefont{T.}~\bibnamefont{Herdtle}} \bibnamefont{and}
  \bibinfo{author}{\bibfnamefont{H.}~\bibnamefont{Aref}}, \bibinfo{journal}{J.
  Fluid Mech.} \textbf{\bibinfo{volume}{241}}, \bibinfo{pages}{233}
  (\bibinfo{year}{1992}).

\bibitem[{\citenamefont{Segel et~al.}(1993)\citenamefont{Segel, Mukamel,
  Krichevsky, and Stavans}}]{Segeletal93}
\bibinfo{author}{\bibfnamefont{D.}~\bibnamefont{Segel}},
  \bibinfo{author}{\bibfnamefont{D.}~\bibnamefont{Mukamel}},
  \bibinfo{author}{\bibfnamefont{O.}~\bibnamefont{Krichevsky}},
  \bibnamefont{and} \bibinfo{author}{\bibfnamefont{J.}~\bibnamefont{Stavans}},
  \bibinfo{journal}{Phys. Rev. E} \textbf{\bibinfo{volume}{47}},
  \bibinfo{pages}{812} (\bibinfo{year}{1993}).

\bibitem[{\citenamefont{Neubert and Schreckenberg}(1997)}]{NeuSch97}
\bibinfo{author}{\bibfnamefont{L.}~\bibnamefont{Neubert}} \bibnamefont{and}
  \bibinfo{author}{\bibfnamefont{M.}~\bibnamefont{Schreckenberg}},
  \bibinfo{journal}{Physica A} \textbf{\bibinfo{volume}{240}},
  \bibinfo{pages}{491} (\bibinfo{year}{1997}).

\bibitem[{\citenamefont{Rutenberg and McCurdy}(2006)}]{Rutenberg05}
\bibinfo{author}{\bibfnamefont{A.~D.} \bibnamefont{Rutenberg}}
  \bibnamefont{and} \bibinfo{author}{\bibfnamefont{M.~B.}
  \bibnamefont{McCurdy}}, \bibinfo{journal}{Phys. Rev. E}
  \textbf{\bibinfo{volume}{73}}, \bibinfo{pages}{011403}
  (\bibinfo{year}{2006}).

\bibitem[{\citenamefont{MacPherson and
  Srolovitz}(2007)}]{MacPhersonSrolovitz2007}
\bibinfo{author}{\bibfnamefont{R.~D.} \bibnamefont{MacPherson}}
  \bibnamefont{and} \bibinfo{author}{\bibfnamefont{D.~J.}
  \bibnamefont{Srolovitz}}, \bibinfo{journal}{Nature}
  \textbf{\bibinfo{volume}{446}}, \bibinfo{pages}{1053} (\bibinfo{year}{2007}).

\bibitem[{\citenamefont{Durian et~al.}(1990)\citenamefont{Durian, Weitz, and
  Pine}}]{DurianWeitzPine90}
\bibinfo{author}{\bibfnamefont{D.~J.} \bibnamefont{Durian}},
  \bibinfo{author}{\bibfnamefont{D.~A.} \bibnamefont{Weitz}}, \bibnamefont{and}
  \bibinfo{author}{\bibfnamefont{D.~J.} \bibnamefont{Pine}},
  \bibinfo{journal}{J. Phys. Condens. Matter} \textbf{\bibinfo{volume}{2}},
  \bibinfo{pages}{433} (\bibinfo{year}{1990}).

\bibitem[{\citenamefont{Durian et~al.}(1991{\natexlab{a}})\citenamefont{Durian,
  Weitz, and Pine}}]{DurianWeitzPine91a}
\bibinfo{author}{\bibfnamefont{D.~J.} \bibnamefont{Durian}},
  \bibinfo{author}{\bibfnamefont{D.~A.} \bibnamefont{Weitz}}, \bibnamefont{and}
  \bibinfo{author}{\bibfnamefont{D.~J.} \bibnamefont{Pine}},
  \bibinfo{journal}{Science} \textbf{\bibinfo{volume}{252}},
  \bibinfo{pages}{686} (\bibinfo{year}{1991}{\natexlab{a}}).

\bibitem[{\citenamefont{Durian et~al.}(1991{\natexlab{b}})\citenamefont{Durian,
  Weitz, and Pine}}]{DurianWeitzPine91b}
\bibinfo{author}{\bibfnamefont{D.~J.} \bibnamefont{Durian}},
  \bibinfo{author}{\bibfnamefont{D.~A.} \bibnamefont{Weitz}}, \bibnamefont{and}
  \bibinfo{author}{\bibfnamefont{D.~J.} \bibnamefont{Pine}},
  \bibinfo{journal}{Phys. Rev. A} \textbf{\bibinfo{volume}{44}},
  \bibinfo{pages}{7902} (\bibinfo{year}{1991}{\natexlab{b}}).

\bibitem[{\citenamefont{Hutzler and Weaire}(2000)}]{HutzlerWeaire00}
\bibinfo{author}{\bibfnamefont{S.}~\bibnamefont{Hutzler}} \bibnamefont{and}
  \bibinfo{author}{\bibfnamefont{D.}~\bibnamefont{Weaire}},
  \bibinfo{journal}{Philos. Mag. Lett.} \textbf{\bibinfo{volume}{80}},
  \bibinfo{pages}{419} (\bibinfo{year}{2000}).

\bibitem[{\citenamefont{Gonatas et~al.}(1995)\citenamefont{Gonatas, Leigh,
  Yodh, Glazier, and Prause}}]{Gonatusetal95}
\bibinfo{author}{\bibfnamefont{C.~P.} \bibnamefont{Gonatas}},
  \bibinfo{author}{\bibfnamefont{J.~S.} \bibnamefont{Leigh}},
  \bibinfo{author}{\bibfnamefont{A.~G.} \bibnamefont{Yodh}},
  \bibinfo{author}{\bibfnamefont{J.~A.} \bibnamefont{Glazier}},
  \bibnamefont{and} \bibinfo{author}{\bibfnamefont{B.}~\bibnamefont{Prause}},
  \bibinfo{journal}{Phys. Rev. Lett.} \textbf{\bibinfo{volume}{75}},
  \bibinfo{pages}{573} (\bibinfo{year}{1995}).

\bibitem[{\citenamefont{Monnereau and Vignes-Adler}(1998)}]{Adler98}
\bibinfo{author}{\bibfnamefont{C.}~\bibnamefont{Monnereau}} \bibnamefont{and}
  \bibinfo{author}{\bibfnamefont{M.}~\bibnamefont{Vignes-Adler}},
  \bibinfo{journal}{Phys. Rev. Lett.} \textbf{\bibinfo{volume}{80}},
  \bibinfo{pages}{5228} (\bibinfo{year}{1998}).

\bibitem[{\citenamefont{Lambert et~al.}(2005)\citenamefont{Lambert, Cantat,
  Delannay, Renault, Graner, Glazier, Veretennikov, and
  Cloetens}}]{GlazierGraner05}
\bibinfo{author}{\bibfnamefont{J.}~\bibnamefont{Lambert}},
  \bibinfo{author}{\bibfnamefont{I.}~\bibnamefont{Cantat}},
  \bibinfo{author}{\bibfnamefont{R.}~\bibnamefont{Delannay}},
  \bibinfo{author}{\bibfnamefont{A.}~\bibnamefont{Renault}},
  \bibinfo{author}{\bibfnamefont{F.}~\bibnamefont{Graner}},
  \bibinfo{author}{\bibfnamefont{J.~A.} \bibnamefont{Glazier}},
  \bibinfo{author}{\bibfnamefont{I.}~\bibnamefont{Veretennikov}},
  \bibnamefont{and} \bibinfo{author}{\bibfnamefont{P.}~\bibnamefont{Cloetens}},
  \bibinfo{journal}{Coll. and Surf. A} \textbf{\bibinfo{volume}{263}},
  \bibinfo{pages}{295} (\bibinfo{year}{2005}).

\bibitem[{\citenamefont{Lambert et~al.}(2010)\citenamefont{Lambert, Mokso,
  Cantat, Cloetens, Glazier, Graner, and Delannay}}]{GlazierGraner10}
\bibinfo{author}{\bibfnamefont{J.}~\bibnamefont{Lambert}},
  \bibinfo{author}{\bibfnamefont{R.}~\bibnamefont{Mokso}},
  \bibinfo{author}{\bibfnamefont{I.}~\bibnamefont{Cantat}},
  \bibinfo{author}{\bibfnamefont{P.}~\bibnamefont{Cloetens}},
  \bibinfo{author}{\bibfnamefont{J.~A.} \bibnamefont{Glazier}},
  \bibinfo{author}{\bibfnamefont{F.}~\bibnamefont{Graner}}, \bibnamefont{and}
  \bibinfo{author}{\bibfnamefont{R.}~\bibnamefont{Delannay}},
  \bibinfo{journal}{Phys. Rev. Lett.} \textbf{\bibinfo{volume}{104}},
  \bibinfo{pages}{248304} (\bibinfo{year}{2010}).

\bibitem[{\citenamefont{Gardiner et~al.}(2000)\citenamefont{Gardiner,
  Dlugogorski, and Jameson}}]{Jameson99}
\bibinfo{author}{\bibfnamefont{B.~S.} \bibnamefont{Gardiner}},
  \bibinfo{author}{\bibfnamefont{B.~Z.} \bibnamefont{Dlugogorski}},
  \bibnamefont{and} \bibinfo{author}{\bibfnamefont{G.~J.}
  \bibnamefont{Jameson}}, \bibinfo{journal}{Philos. Mag. A}
  \textbf{\bibinfo{volume}{80}}, \bibinfo{pages}{981} (\bibinfo{year}{2000}).

\bibitem[{\citenamefont{Hilgenfeldt
  et~al.}(2001{\natexlab{a}})\citenamefont{Hilgenfeldt, Koehler, and
  Stone}}]{StoneKoehlerHilgenfeldt01}
\bibinfo{author}{\bibfnamefont{S.}~\bibnamefont{Hilgenfeldt}},
  \bibinfo{author}{\bibfnamefont{S.~A.} \bibnamefont{Koehler}},
  \bibnamefont{and} \bibinfo{author}{\bibfnamefont{H.~A.} \bibnamefont{Stone}},
  \bibinfo{journal}{Phys. Rev. Lett.} \textbf{\bibinfo{volume}{86}},
  \bibinfo{pages}{4704} (\bibinfo{year}{2001}{\natexlab{a}}).

\bibitem[{\citenamefont{Vera and Durian}(2002)}]{VeraDurian02}
\bibinfo{author}{\bibfnamefont{M.~U.} \bibnamefont{Vera}} \bibnamefont{and}
  \bibinfo{author}{\bibfnamefont{D.~J.} \bibnamefont{Durian}},
  \bibinfo{journal}{Phys. Rev. Lett.} \textbf{\bibinfo{volume}{88}},
  \bibinfo{pages}{088304} (\bibinfo{year}{2002}).

\bibitem[{\citenamefont{Ga$\tilde{\textrm{n}}$\'{a}n-Calvo
  et~al.}(2004)\citenamefont{Ga$\tilde{\textrm{n}}$\'{a}n-Calvo, Fernandez,
  Marquez~Oliver, and Marquez}}]{Marquezetal04}
\bibinfo{author}{\bibfnamefont{A.~M.}
  \bibnamefont{Ga$\tilde{\textrm{n}}$\'{a}n-Calvo}},
  \bibinfo{author}{\bibfnamefont{J.~M.} \bibnamefont{Fernandez}},
  \bibinfo{author}{\bibfnamefont{A.}~\bibnamefont{Marquez~Oliver}},
  \bibnamefont{and} \bibinfo{author}{\bibfnamefont{M.}~\bibnamefont{Marquez}},
  \bibinfo{journal}{Appl. Phys. Lett.} \textbf{\bibinfo{volume}{84}},
  \bibinfo{pages}{4989} (\bibinfo{year}{2004}).

\bibitem[{\citenamefont{Feitosa and Durian}(2008)}]{FeitosaDurian08}
\bibinfo{author}{\bibfnamefont{K.}~\bibnamefont{Feitosa}} \bibnamefont{and}
  \bibinfo{author}{\bibfnamefont{D.~J.} \bibnamefont{Durian}},
  \bibinfo{journal}{Eur. Phys. J. E} \textbf{\bibinfo{volume}{26}},
  \bibinfo{pages}{309} (\bibinfo{year}{2008}).

\bibitem[{\citenamefont{Bolton and Weaire}(1991)}]{BoltonWeaire91}
\bibinfo{author}{\bibfnamefont{F.}~\bibnamefont{Bolton}} \bibnamefont{and}
  \bibinfo{author}{\bibfnamefont{D.}~\bibnamefont{Weaire}},
  \bibinfo{journal}{Phil. Mag. B} \textbf{\bibinfo{volume}{63}},
  \bibinfo{pages}{795} (\bibinfo{year}{1991}).

\bibitem[{\citenamefont{Weaire}(1999)}]{Weaire99}
\bibinfo{author}{\bibfnamefont{D.}~\bibnamefont{Weaire}},
  \bibinfo{journal}{Philos. Mag. Lett.} \textbf{\bibinfo{volume}{79}},
  \bibinfo{pages}{491} (\bibinfo{year}{1999}).

\bibitem[{\citenamefont{Teixeira and Fortes}(2005)}]{Fortes05}
\bibinfo{author}{\bibfnamefont{P.~I.~C.} \bibnamefont{Teixeira}}
  \bibnamefont{and} \bibinfo{author}{\bibfnamefont{M.~A.}
  \bibnamefont{Fortes}}, \bibinfo{journal}{Philos Mag.}
  \textbf{\bibinfo{volume}{85}}, \bibinfo{pages}{1303} (\bibinfo{year}{2005}).

\bibitem[{\citenamefont{Mancini and Oguey}(2007)}]{Mancini07}
\bibinfo{author}{\bibfnamefont{M.}~\bibnamefont{Mancini}} \bibnamefont{and}
  \bibinfo{author}{\bibfnamefont{C.}~\bibnamefont{Oguey}},
  \bibinfo{journal}{Eur. Phys. J. E} \textbf{\bibinfo{volume}{22}},
  \bibinfo{pages}{181} (\bibinfo{year}{2007}).

\bibitem[{\citenamefont{Bolton and Weaire}(1992)}]{BoltonWeaire92}
\bibinfo{author}{\bibfnamefont{F.}~\bibnamefont{Bolton}} \bibnamefont{and}
  \bibinfo{author}{\bibfnamefont{D.}~\bibnamefont{Weaire}},
  \bibinfo{journal}{Philos. Mag. B} \textbf{\bibinfo{volume}{65}},
  \bibinfo{pages}{473} (\bibinfo{year}{1992}).

\bibitem[{\citenamefont{Fortuna et~al.}(2012)\citenamefont{Fortuna, Thomas,
  de~Almeida, and Graner}}]{Graneretalarxiv}
\bibinfo{author}{\bibfnamefont{I.}~\bibnamefont{Fortuna}},
  \bibinfo{author}{\bibfnamefont{G.~L.} \bibnamefont{Thomas}},
  \bibinfo{author}{\bibfnamefont{R.~M.~C.} \bibnamefont{de~Almeida}},
  \bibnamefont{and} \bibinfo{author}{\bibfnamefont{F.}~\bibnamefont{Graner}},
  \bibinfo{journal}{Phys. Rev. Lett.} \textbf{\bibinfo{volume}{108}},
  \bibinfo{pages}{248301} (\bibinfo{year}{2012}).

\bibitem[{\citenamefont{Marchalot et~al.}(2008)\citenamefont{Marchalot,
  Lambert, Cantat, Tabeling, and Jullien}}]{Marchalot08}
\bibinfo{author}{\bibfnamefont{J.}~\bibnamefont{Marchalot}},
  \bibinfo{author}{\bibfnamefont{J.}~\bibnamefont{Lambert}},
  \bibinfo{author}{\bibfnamefont{I.}~\bibnamefont{Cantat}},
  \bibinfo{author}{\bibfnamefont{P.}~\bibnamefont{Tabeling}}, \bibnamefont{and}
  \bibinfo{author}{\bibfnamefont{M.-C.} \bibnamefont{Jullien}},
  \bibinfo{journal}{Europhys. Lett} \textbf{\bibinfo{volume}{83}},
  \bibinfo{pages}{64006} (\bibinfo{year}{2008}).

\bibitem[{\citenamefont{Quilliet et~al.}(2008)\citenamefont{Quilliet,
  Ataei~Talebi, Rabaud, K\"afer, Cox, and Graner}}]{Quilliet2008}
\bibinfo{author}{\bibfnamefont{C.}~\bibnamefont{Quilliet}},
  \bibinfo{author}{\bibfnamefont{S.}~\bibnamefont{Ataei~Talebi}},
  \bibinfo{author}{\bibfnamefont{D.}~\bibnamefont{Rabaud}},
  \bibinfo{author}{\bibfnamefont{J.}~\bibnamefont{K\"afer}},
  \bibinfo{author}{\bibfnamefont{S.~J.} \bibnamefont{Cox}}, \bibnamefont{and}
  \bibinfo{author}{\bibfnamefont{F.}~\bibnamefont{Graner}},
  \bibinfo{journal}{Philos. Mag. Lett.} \textbf{\bibinfo{volume}{88}},
  \bibinfo{pages}{651} (\bibinfo{year}{2008}).

\bibitem[{\citenamefont{Duplat et~al.}(2011)\citenamefont{Duplat, Bossa, and
  Villermaux}}]{Duplat2011}
\bibinfo{author}{\bibfnamefont{J.}~\bibnamefont{Duplat}},
  \bibinfo{author}{\bibfnamefont{B.}~\bibnamefont{Bossa}}, \bibnamefont{and}
  \bibinfo{author}{\bibfnamefont{E.}~\bibnamefont{Villermaux}},
  \bibinfo{journal}{J. Fluid Mech.} \textbf{\bibinfo{volume}{673}},
  \bibinfo{pages}{147} (\bibinfo{year}{2011}).

\bibitem[{\citenamefont{Flyvbjerg}(1993)}]{Flyvbjerg93}
\bibinfo{author}{\bibfnamefont{H.}~\bibnamefont{Flyvbjerg}},
  \bibinfo{journal}{Phys. Rev. E} \textbf{\bibinfo{volume}{47}},
  \bibinfo{pages}{4037} (\bibinfo{year}{1993}).

\bibitem[{\citenamefont{Durand et~al.}(2011)\citenamefont{Durand,
  K$\ddot{\textrm{a}}$fer, Quilliet, Cox, Talebi, and Graner}}]{Graneretal11}
\bibinfo{author}{\bibfnamefont{M.}~\bibnamefont{Durand}},
  \bibinfo{author}{\bibfnamefont{J.}~\bibnamefont{K$\ddot{\textrm{a}}$fer}},
  \bibinfo{author}{\bibfnamefont{C.}~\bibnamefont{Quilliet}},
  \bibinfo{author}{\bibfnamefont{S.}~\bibnamefont{Cox}},
  \bibinfo{author}{\bibfnamefont{S.~A.} \bibnamefont{Talebi}},
  \bibnamefont{and} \bibinfo{author}{\bibfnamefont{F.}~\bibnamefont{Graner}},
  \bibinfo{journal}{Phys. Rev. Lett.} \textbf{\bibinfo{volume}{107}},
  \bibinfo{pages}{168304} (\bibinfo{year}{2011}).

\bibitem[{\citenamefont{Chiu}(1995)}]{ChiuReview}
\bibinfo{author}{\bibfnamefont{S.~N.} \bibnamefont{Chiu}},
  \bibinfo{journal}{Materials Characterization} \textbf{\bibinfo{volume}{34}},
  \bibinfo{pages}{149} (\bibinfo{year}{1995}).

\bibitem[{\citenamefont{Lewis}(1928)}]{Lewis1928}
\bibinfo{author}{\bibfnamefont{F.~T.} \bibnamefont{Lewis}},
  \bibinfo{journal}{Anat. Rec.} \textbf{\bibinfo{volume}{38}},
  \bibinfo{pages}{341} (\bibinfo{year}{1928}).

\bibitem[{\citenamefont{Lewis}(1930)}]{Lewis1930}
\bibinfo{author}{\bibfnamefont{F.~T.} \bibnamefont{Lewis}},
  \bibinfo{journal}{Anat. Rec.} \textbf{\bibinfo{volume}{47}},
  \bibinfo{pages}{59} (\bibinfo{year}{1930}).

\bibitem[{\citenamefont{Rivier and Lissowski}(1982)}]{RivierLissowski82}
\bibinfo{author}{\bibfnamefont{N.}~\bibnamefont{Rivier}} \bibnamefont{and}
  \bibinfo{author}{\bibfnamefont{A.}~\bibnamefont{Lissowski}},
  \bibinfo{journal}{J. Phys. A: Math. Gen.} \textbf{\bibinfo{volume}{15}},
  \bibinfo{pages}{143} (\bibinfo{year}{1982}).

\bibitem[{\citenamefont{Rivier}(1985)}]{Rivier85}
\bibinfo{author}{\bibfnamefont{N.}~\bibnamefont{Rivier}},
  \bibinfo{journal}{Phil. Mag. B} \textbf{\bibinfo{volume}{52}},
  \bibinfo{pages}{795} (\bibinfo{year}{1985}).

\bibitem[{\citenamefont{Szeto and Tam}(1995)}]{SzetoTam95}
\bibinfo{author}{\bibfnamefont{K.~Y.} \bibnamefont{Szeto}} \bibnamefont{and}
  \bibinfo{author}{\bibfnamefont{W.~Y.} \bibnamefont{Tam}},
  \bibinfo{journal}{Physica A} \textbf{\bibinfo{volume}{221}},
  \bibinfo{pages}{256} (\bibinfo{year}{1995}).

\bibitem[{\citenamefont{Saraiva et~al.}(2009)\citenamefont{Saraiva, Pina,
  Bandeira, and Antunes}}]{Saraivaetal09}
\bibinfo{author}{\bibfnamefont{J.}~\bibnamefont{Saraiva}},
  \bibinfo{author}{\bibfnamefont{P.}~\bibnamefont{Pina}},
  \bibinfo{author}{\bibfnamefont{L.}~\bibnamefont{Bandeira}}, \bibnamefont{and}
  \bibinfo{author}{\bibfnamefont{J.}~\bibnamefont{Antunes}},
  \bibinfo{journal}{Philos. Mag. Lett.} \textbf{\bibinfo{volume}{89}},
  \bibinfo{pages}{185} (\bibinfo{year}{2009}).

\bibitem[{\citenamefont{Lambert et~al.}(2012)\citenamefont{Lambert, Graner,
  Delannay, and Cantat}}]{Lambert2012}
\bibinfo{author}{\bibfnamefont{J.}~\bibnamefont{Lambert}},
  \bibinfo{author}{\bibfnamefont{F.}~\bibnamefont{Graner}},
  \bibinfo{author}{\bibfnamefont{R.}~\bibnamefont{Delannay}}, \bibnamefont{and}
  \bibinfo{author}{\bibfnamefont{I.}~\bibnamefont{Cantat}},
  \bibinfo{journal}{Europhys. Lett.} \textbf{\bibinfo{volume}{99}},
  \bibinfo{pages}{48003} (\bibinfo{year}{2012}).

\bibitem[{\citenamefont{Newhall et~al.}(2012)\citenamefont{Newhall, Pontani,
  Jorjadze, Hilgenfeldt, and Bruji\'c}}]{BrujicPRL12}
\bibinfo{author}{\bibfnamefont{K.~A.} \bibnamefont{Newhall}},
  \bibinfo{author}{\bibfnamefont{L.~L.} \bibnamefont{Pontani}},
  \bibinfo{author}{\bibfnamefont{I.}~\bibnamefont{Jorjadze}},
  \bibinfo{author}{\bibfnamefont{S.}~\bibnamefont{Hilgenfeldt}},
  \bibnamefont{and} \bibinfo{author}{\bibfnamefont{J.}~\bibnamefont{Bruji\'c}},
  \bibinfo{journal}{Phys. Rev. Lett.} \textbf{\bibinfo{volume}{108}},
  \bibinfo{pages}{268001} (\bibinfo{year}{2012}).

\bibitem[{\citenamefont{Clusel et~al.}(2009)\citenamefont{Clusel, Corwin,
  Siemens, and Bruji\'c}}]{BrujicNature09}
\bibinfo{author}{\bibfnamefont{M.}~\bibnamefont{Clusel}},
  \bibinfo{author}{\bibfnamefont{E.~I.} \bibnamefont{Corwin}},
  \bibinfo{author}{\bibfnamefont{A.~O.~N.} \bibnamefont{Siemens}},
  \bibnamefont{and} \bibinfo{author}{\bibfnamefont{J.}~\bibnamefont{Bruji\'c}},
  \bibinfo{journal}{Nature} \textbf{\bibinfo{volume}{460}},
  \bibinfo{pages}{611} (\bibinfo{year}{2009}).

\bibitem[{\citenamefont{Graner et~al.}(2000)\citenamefont{Graner, Jiang,
  Janiaud, and Flament}}]{Graner2000}
\bibinfo{author}{\bibfnamefont{F.}~\bibnamefont{Graner}},
  \bibinfo{author}{\bibfnamefont{Y.}~\bibnamefont{Jiang}},
  \bibinfo{author}{\bibfnamefont{E.}~\bibnamefont{Janiaud}}, \bibnamefont{and}
  \bibinfo{author}{\bibfnamefont{C.}~\bibnamefont{Flament}},
  \bibinfo{journal}{Phys. Rev. E} \textbf{\bibinfo{volume}{63}},
  \bibinfo{pages}{011402} (\bibinfo{year}{2000}).

\bibitem[{\citenamefont{Hilgenfeldt
  et~al.}(2001{\natexlab{b}})\citenamefont{Hilgenfeldt, Kraynik, Koehler, and
  Stone}}]{HilgenfeldtKraynikKoehlerStone00}
\bibinfo{author}{\bibfnamefont{S.}~\bibnamefont{Hilgenfeldt}},
  \bibinfo{author}{\bibfnamefont{A.~M.} \bibnamefont{Kraynik}},
  \bibinfo{author}{\bibfnamefont{S.~A.} \bibnamefont{Koehler}},
  \bibnamefont{and} \bibinfo{author}{\bibfnamefont{H.~A.} \bibnamefont{Stone}},
  \bibinfo{journal}{Phys. Rev. Lett.} \textbf{\bibinfo{volume}{86}},
  \bibinfo{pages}{2685} (\bibinfo{year}{2001}{\natexlab{b}}).

\bibitem[{\citenamefont{Hilgenfeldt et~al.}(2004)\citenamefont{Hilgenfeldt,
  Kraynik, Reinelt, and Sullivan}}]{HilgenfeldtKraynik04}
\bibinfo{author}{\bibfnamefont{S.}~\bibnamefont{Hilgenfeldt}},
  \bibinfo{author}{\bibfnamefont{A.~M.} \bibnamefont{Kraynik}},
  \bibinfo{author}{\bibfnamefont{D.~A.} \bibnamefont{Reinelt}},
  \bibnamefont{and} \bibinfo{author}{\bibfnamefont{J.}~\bibnamefont{Sullivan}},
  \bibinfo{journal}{Europhys. Lett.} \textbf{\bibinfo{volume}{67}},
  \bibinfo{pages}{484} (\bibinfo{year}{2004}).

\bibitem[{\citenamefont{Glicksman}(2005)}]{Glicksman05}
\bibinfo{author}{\bibfnamefont{M.~E.} \bibnamefont{Glicksman}},
  \bibinfo{journal}{Philos. Mag.} \textbf{\bibinfo{volume}{85}},
  \bibinfo{pages}{3} (\bibinfo{year}{2005}).

\bibitem[{\citenamefont{Glicksman et~al.}(2007)\citenamefont{Glicksman, Rios,
  and Lewis}}]{GlicksmanRiosLewis07}
\bibinfo{author}{\bibfnamefont{M.~E.} \bibnamefont{Glicksman}},
  \bibinfo{author}{\bibfnamefont{P.~R.} \bibnamefont{Rios}}, \bibnamefont{and}
  \bibinfo{author}{\bibfnamefont{D.~J.} \bibnamefont{Lewis}},
  \bibinfo{journal}{Acta Materialia} \textbf{\bibinfo{volume}{55}},
  \bibinfo{pages}{4167} (\bibinfo{year}{2007}).

\end{thebibliography}

\end{document}